\documentclass[reprint,amsmath,amssymb,aps]{revtex4-1}
\usepackage{graphicx}
\graphicspath{{Figures/}}

\usepackage{dcolumn}
\usepackage{bm}

\usepackage{amssymb}
\usepackage{amsmath}
\usepackage{epsfig}
\usepackage{chemarr}
\usepackage{url}
\usepackage{natbib}
\usepackage{color}

\usepackage[final]{changes}
\definechangesauthor[name={Nachi's remarks}, color=orange]{Nachi}

\begin{document}

\title{Supervised Learning in Physical Networks: \\ From Machine Learning to Learning Machines}

\author{Menachem Stern${}^1$, Daniel Hexner${}^2$, Jason W. Rocks${}^3$ and Andrea J. Liu${}^1$}

\affiliation{${}^1$ Department of Physics and Astronomy, University of Pennsylvania, Philadelphia, PA 19104}

\affiliation{${}^2$ Department of Mechanical Engineering, Technion, Haifa, 32000, Israel}

\affiliation{${}^3$ Department of Physics, Boston University, Boston, MA 02139}


\date{\today}

\begin{abstract}


Materials and machines are often designed with particular goals in mind, so that they exhibit desired responses to given forces or constraints. Here we explore an alternative approach, namely physical \textit{coupled learning}. In this paradigm, the system is not initially designed to accomplish a task, but physically adapts to applied forces to develop the ability to perform the task. Crucially, we require coupled learning to be facilitated by physically plausible learning rules, meaning that learning requires only local responses and no explicit information about the desired functionality.
We show that such local learning rules can be derived for any physical network, whether in equilibrium or in steady state, with specific focus on two particular systems, namely disordered flow networks and elastic networks.  By applying and adapting advances of statistical learning theory to the physical world, we demonstrate the plausibility of new classes of smart metamaterials capable of adapting to users' needs \it{in-situ}.


\end{abstract}

\pacs{Valid PACS appear here}
\maketitle

\section{Introduction}

Engineered materials are typically designed to have particular sets of properties or functions~\cite{norman2013design}. The design process often involves numerous trial and error iterations, during which the system is repeatedly tested for the desired functionality~\cite{thomke1998managing}, modified and tested again. Alternatively, rational design processes start from detailed knowledge of material components and typically use computation to predict the consequences of tweaking the system to sift through many possibilities.

A second class of strategies is based on \emph{learning}, where systems can adjust or be adjusted at the microscopic scale, in response to training examples, to develop the desired functionality. Until recently, learning strategies were primarily restricted to nonphysical networks such as neural networks on a computer. One class of methods, which we will refer to as ``global supervised learning," is ubiquitous for problems such as data classification~\cite{kotsiantis2007supervised,mehta2019high}. These methods are based on the optimization of an objective, cost of loss function that is ``global" in that it depends on all of the microscopic details of the system. 
In the context of neural networks, for example, global supervised learning optimizes the network by modifying a set of learning degrees of freedom ({\it e.g.} weights) controlling the signal propagation between the input and output sections of the network. 

Global supervised learning was recently used to design flow and elastic networks--actual physical systems--with particular desired functions, e.g. allostery~\cite{rocks2017designing,rocks2019limits}. In this physical context, such learning methods were dubbed \textit{tuning}, since modifying the learning degrees of freedom (e.g. pipe conductances or spring constants in flow or mechanical networks, respectively) requires external intervention at the microscopic level.

In contrast, natural systems such as the brain evolve to obtain desired functions using a fundamentally different framework of learning, which we will refer to as ``local learning." Crucially, this evolution is entirely autonomous, requiring no external designer for evaluation of the current state of the system and its subsequent modification. In local learning, parts of the network can only adapt due to the local information available in their immediate vicinity (e.g. a synapse adapts in response to the activities only of the neurons directly connected to it~\cite{richards2019dendritic}). It is particularly useful to apply this learning approach in physical networks such as flow or mechanical networks because the microscopic elements of such networks cannot perform computations and do not encode information about the desired functionality \textit{a priori}. 

Training physical networks for desired function, using either global or local supervised learning, involves two different sets of degrees of freedom. First, networks respond to source constraints by minimizing a scalar function with respect to their \textit{physical degrees of freedom}. For example, central-force spring networks minimize the elastic energy by adjusting the positions of their nodes to achieve force balance on every node, while flow networks minimize the power dissipation by adjusting the currents on edges to obey Kirchhoff's law at every node. Second, networks can learn specific desired target responses to sources by adjusting their \textit{learning degrees of freedom}. These degrees of freedom could correspond to the stiffnesses or equilibrium lengths of springs in a mechanical network or the conductances of edges in a flow network. In global supervised learning, these degrees of freedom are adjusted to optimize the cost function.

We will consider the case where the desired outcome of the learning process is to achieve tailored responses of ``target" edges or nodes to external constraints applied at ``source" edges or nodes. 
For example, an allosteric response in a mechanical network corresponds to a desired strain at a set of target edges in response to a strain applied at source edges. Similarly, for the simplest ``flow allosteric" response, a pressure drop across a source edge in a flow network leads to a desired pressure drop across a target edge elsewhere in the network~\cite{rocks2019revealing}. 

Recently it was shown that a local supervised learning process of ``directed aging," in which the time evolution of a disordered system is driven by applied stresses, can be used to create mechanical metamaterials with desired properties or functions~\cite{pashine2019directed,hexner2019effect,hexner2019periodic}. For example, Refs.~\cite{hexner2019effect,hexner2019periodic} consider a form of directed aging for a mechanical spring network in which the learning degrees of freedom are the equilibrium lengths of the springs comprising the network edges. The equilibrium length of each spring lengthens/shortens if the spring is placed under extension/compression--this is a local response to local extension/compression. Isotropic compression~\cite{pashine2019directed,hexner2019effect} or repeated cycles of isotropic compression and expansion can drive the Poisson's ratio of such a network from positive to negative values, while cycles of compression/stretching oscillations of source and target edges can lead to allosteric response~\cite{hexner2019periodic}.

Similarly, local rules in growing vascular networks~\cite{ronellenfitsch2016global} and folding sheets~\cite{stern2020supervised,stern2018shaping} allow those systems to learn properties or functions autonomously. The great advantage of local supervised learning such as directed aging over global supervised learning is that the process is scalable--it can be applied to train large systems without having to manually modify their parts~\cite{pashine2019directed}. In addition, directed aging does not require detailed knowledge of microscopic interactions or the ability to manipulate (microscopic) learning degrees of freedom~\cite{pashine2019directed}. 

While directed aging methods are successful in training certain physical networks for desired functions, they fail in others, particularly in highly over-constrained networks such as flow networks and high-coordination mechanical networks. Failure occurs because directed aging minimizes the physical cost function of a desired state of the network, rather than the cost function whose minimization corresponds directly to the desired function in global supervised learning. 

Here we propose a general framework, which we call ``coupled local supervised learning," for physical networks such as flow or mechanical networks. 
The rules are designed to adjust learning degrees of freedom in response to local conditions such as the tension on a spring in a mechanical network or the current through a pipe of a flow network. The framework provides a way of deriving rules that lead to modifications of learning degrees of freedom that are extremely similar to those obtained by minimizing the cost function. As a result, they are as likely to succeed in obtaining the desired response as global supervised learning would be.

The coupled learning framework is inspired by advances in neuroscience and computer science~\cite{movellan1991contrastive,baldi2016theory,bengio2015early,scellier2017equilibrium,bartunov2018assessing} known as ``contrastive Hebbian learning." As in contrastive Hebbian learning, one considers the response in two steady states of the system, one in which only source constraints are applied (\textit{free state}), the other where source and target constraints are applied simultaneously (\textit{clamped state}). The particular rules we introduce, which we call \textit{coupled learning} rules, are also inspired by the strategy of ``equilibrium propagation" \cite{scellier2017equilibrium}, which promotes infinitesimal nudging and hence close proximity between the free and clamped states.  
In part II, we first show in detail how coupled learning works for flow networks (Sec. A). We successfully train such networks to obtain complex functionalities. We then discuss the general framework of coupled learning in generic physical networks (Sec. B) and 
and apply these ideas to non-linear elastic networks (Sec. C). We demonstrate our learning framework on a standard classification problem, distinguishing handwritten digits (Sec. D).

It is important to note that implementation of coupled learning should be possible in real systems. In part III we therefore consider complications that may arise in realistic learning scenarios. We first derive an approximate version of the local learning rule that may be more easily implemented experimentally in a flow network (Sec. A). While inferior to the full learning rule, approximate coupled learning still gives rise to desired functionality. We then discuss limitations due to noisy measurements of the physical degrees of freedom, that limit the usefulness of small nudging, and the implications of drifting in the learning degrees of freedom (Sec. B). Finally, we address the effect of networks size on the physical relaxation time and the coupled learning rules (Sec. C).

Following the introduction of the coupled learning rules, in part IV we compare coupled learning to standard global supervised learning frameworks that minimize of a cost function~\cite{rocks2017designing,rocks2019limits,kim2019conformational,ruiz2019tuning}, discussing the experimental realizability of coupled learning in contrast with such methods. We hope this work will stimulate further interest in physically inspired learning, opening possibilities for new classes of metamaterials and machines, able to autonomously evolve and adapt in desired ways due to external influences.




\begin{figure*}
\includegraphics[width=0.98\linewidth]{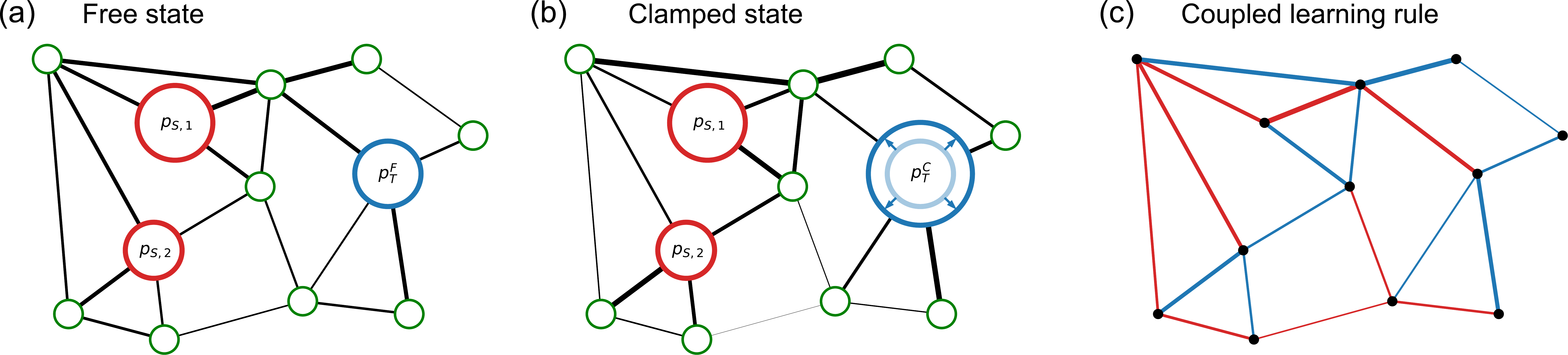}
\caption{Coupled learning in flow networks. a) In the free phase, node pressures are constrained such that source nodes (red) have specific pressure values $\bm{p}_S$. The target node pressures $\bm{p}_T$ and the dissipated power at all pipes $\boldmath{\mathcal{P}}_j$ attain their steady state values due to the natural flow processes in the network. b) In the clamped phase, a supervisor constrains the target node pressures so that they are slightly closer to their desired values compared to the free state. c) A strictly local learning rule, i.e. the modification of pipe conductance according to its response, is proportional to the difference in dissipated power between the free and clamped states. The learning rule thus couples the pressures at the source and target nodes.
\label{fig:Schematic}}
\end{figure*}


\section{Coupled Learning framework}

\subsection{Coupled learning in flow networks}


We first discuss coupled learning in the context of flow networks. Previously, such networks were trained to exhibit the particular function of flow allostery~\cite{rocks2017designing,rocks2019limits} using global supervised learning by minimizing a global cost function. Specifically, the networks were trained to have a desired pressure drop across a target edge (or many target edges) in response to a pressure drop applied across a source edge elsewhere in the network. Here we show how a strictly local 
learning rule can similarly train flow networks.

A flow network is defined by a set of $N$ nodes labeled $\mu$, each carrying a pressure value $p_\mu$; these pressures are the \emph{physical degrees of freedom} of the network. The nodes are connected by pipes $j$ characterized by their conductances $k_j$; these conductances are the \emph{learning degrees of freedom} because modifying the pipe conductances will enable the network to develop the desired function. Assuming each pipe is directed from one node $\mu_1$ to the other $\mu_2$, the flow current in the pipe is given by $I_j=k_j(p_{\mu_1}-p_{\mu_2})\equiv k_j\Delta p_j$. If boundary conditions are applied to the network (e.g. fixed pressure values at some nodes) the network finds a flow steady state, in which the total dissipated power 
\begin{equation}
 \mathcal{P}=\frac{1}{2} \sum_j k_j\Delta p_j^2   
\end{equation} 
is minimized by varying the pressures at unconstrained nodes.

We now define a task for the network to learn as follows. We subdivide the physical degrees of freedom (node pressures) $\{p_\mu\}$ into three types, corresponding to source nodes $p_S$, target nodes $p_T$, and the remaining ``hidden" nodes $p_H$. The desired task is defined such that the target node pressure values reach a set of desired values $\{P_T\}$ when the source node pressures are constrained to the values $\{P_S\}$. A generic disordered flow network does not possess this function, so design strategies are needed to identify appropriate values for the pipe conductance values $\{k_j\}$ that achieve the desired task.

For the network to learn, or adapt autonomously to source pressure constraints, we allow the learning degrees of freedom $\{k_j\}$ (pipe conductances) to vary depending on the physical state of the network $\{p_\mu\}$.  A \textit{learning rule} is an equation of motion for the learning degrees of freedom, taking the form 

\begin{equation}
\begin{aligned}
\dot{k}_j=f(p_\mu;k_j).
\end{aligned}
  \label{eq:LearningRuleDefinition}
\end{equation}

We will focus on local learning rules, where the learning degree of freedom $k_j$ in each pipe $j$ can only change in response to the physical variables (pressure values) $\{p_\mu\}$ on the two nodes associated with pipe $j$, $p_{\mu(j)}$.

We now introduce the framework of \textit{coupled learning}. Let us define two sets of constraints on the pressure values of the network. The \textit{free} state $p_\mu^{F}$ is defined as the state where only the source nodes $p_S$ are constrained to their values $P_S$, allowing $p_T,p_H$ to obtain their steady state (\textit{i.\,e.~} to reach values that minimize the physical cost function, the dissipated power $\mathcal{P}(p_\mu;k_j)$) (Fig.~\ref{fig:Schematic}a). The \textit{clamped} state $p_\mu^{C}$ is the state where both the source and target node pressures $p_S,p_T$ are constrained to $P_S$ and $p_T^C$, respectively, so that only the remaining (hidden) nodes $p_H$ are allowed to change to minimize the dissipated power. The values of the dissipated power in the resulting steady states are denoted $\mathcal{P}^{F}(P_S,p_T^{F},p_H^{F};k_j)$ and $\mathcal{P}^{C}(P_S,p_T^C,p_H^{C};k_j)$ for the free and clamped states, respectively. In the following we simplify notation by suppressing the variables in the parentheses.

A disordered flow network constrained only at its sources (free state) will generally find a steady state in which the target pressures $\{p_T^{F}\}$ differ significantly from the desired responses $\{P_T\}$. To concretely define the clamped state, we introduce a trainer (supervisor) that reads the free state target node pressures and nudges them slightly away from their free state $p_T^{F}$ values by clamping them at
\begin{equation}
 p_T^{C}=p_T^{F} + \eta [P_T - p_T^{F}] \\
 \label{eq:Clampeddef1}
\end{equation}
where $\eta \ll 1$ (Fig.~\ref{fig:Schematic}b). The supervisor imposes pressures on the target nodes that are a small step closer to the desired response $P_T$. We then propose a learning rule for the pipe conductance values (Fig.~\ref{fig:Schematic}c):

\begin{equation}
\begin{aligned}
\dot{k}_j &= \alpha \eta^{-1} \frac{\partial}{\partial k_j} \{\mathcal{P}^{F} - \mathcal{P}^{C}\}= \\
             &= \frac{1}{2}\alpha \eta^{-1} \{[\Delta p^{F}_j]^2 - [\Delta p^{C}_j]^2\},
\end{aligned}
  \label{eq:FlowNetworkRule}
\end{equation}
where $\alpha$ is a scalar learning rate. Note that the derivative of the physically minimized function (power dissipation $\mathcal{P}$) is taken only directly with respect to the \emph{learning} degrees of freedom $\{k_j\}$ (pipe conductances), excluding derivatives of the physical variables with respect to them (e.g. $\frac{\partial p_i}{\partial k_j}$). In the limit of small nudges $\eta\rightarrow 0$ such derivatives cancel out when the difference between the free and clamped terms is taken~\cite{scellier2017equilibrium}.

We note that the learning rule of Eq.~\ref{eq:FlowNetworkRule} is similar to the one derived for analog resistor networks by Kendall et. al.~\cite{kendall2020training}. While their equation has the same form, their definition of the clamped state is different, as they draw currents into the output nodes rather than fixing the output pressures. Their nudging is thus akin to applying a force on the output nodes. The simplest implementation of the learning rule is to iteratively apply Eq.~\ref{eq:FlowNetworkRule} for some period of time between each nudge of the clamped boundary condition in Eq.~\ref{eq:Clampeddef1}. Here we focus on learning in the quasistatic limit where we completely relax the node pressure values to their steady state at each iteration to reach the minimum of the power dissipation $\mathcal{P}$. 

The learning rule of Eq.~\ref{eq:FlowNetworkRule} is manifestly local, as the conductance of a pipe $k_j$ changes only due to the flow through that pipe. Such local learning rules may conceivably be implemented in physical pipes for which the conductance (radius) of the pipe is controlled by the current in it. Note that the network is not required to encode information about the desired state function {\it a priori}. That information is fed into the network by the actions of the external supervisor, which slightly nudges the target node pressures towards the desired state at every iteration. We later show how these properties of coupled learning stand in contrast to tuning algorithms based on optimization of global cost functions.

The intuition underlying coupled learning is straightforward. We wish to obtain a network with the desired target pressures $\{P_T\}$ when source pressures $\{P_S\}$ are applied. We divide the training process into small steps, each of which depends on free and clamped states. At every iteration of Eq.~\ref{eq:Clampeddef1}, the clamped state is initially slightly ``better" than the free state, in that the node pressures at the targets, $\{p_T\}$, are slightly closer to the desired pressures $\{P_T\}$ due to the applied nudge. The learning rule changes the pipe conductances $\{k_j\}$ by adding the gradient at the free state (raising the free state power dissipation) and subtracting the gradient at the clamped state (lowering the clamped state power dissipation). By adjusting the pipe conductances, the network response adapts to more closely reflect the clamped state. After many iterations of the learning rule in Eq.~\ref{eq:FlowNetworkRule}, $\{\dot k_j\}$ approach zero and the free state approaches the desired target pressures $P_T$. By iterating in this fashion, the network carves out a valley in the landscape of the dissipated power $\mathcal{P}$ with respect to the target node pressures $\{p_T\}$ that deepens as one approaches the desired target pressures--much as directed aging does when it is successful~\cite{hexner2019periodic}. This intuition implies that the nudge magnitude value should be small $\eta\ll 1$, as was proposed for equilibrium propagation~\cite{scellier2017equilibrium} (see Appendix A). Such choice allows the gradual modification of the power landscape $\mathcal{P}$ so that it becomes minimal along a valley at the desired state. The approach of using small nudges has been shown to achieve excellent results on hard computational learning problems such as CIFAR-10~\cite{laborieux2020scaling}.

\begin{figure}[!ht]
\includegraphics[width=1.0\linewidth]{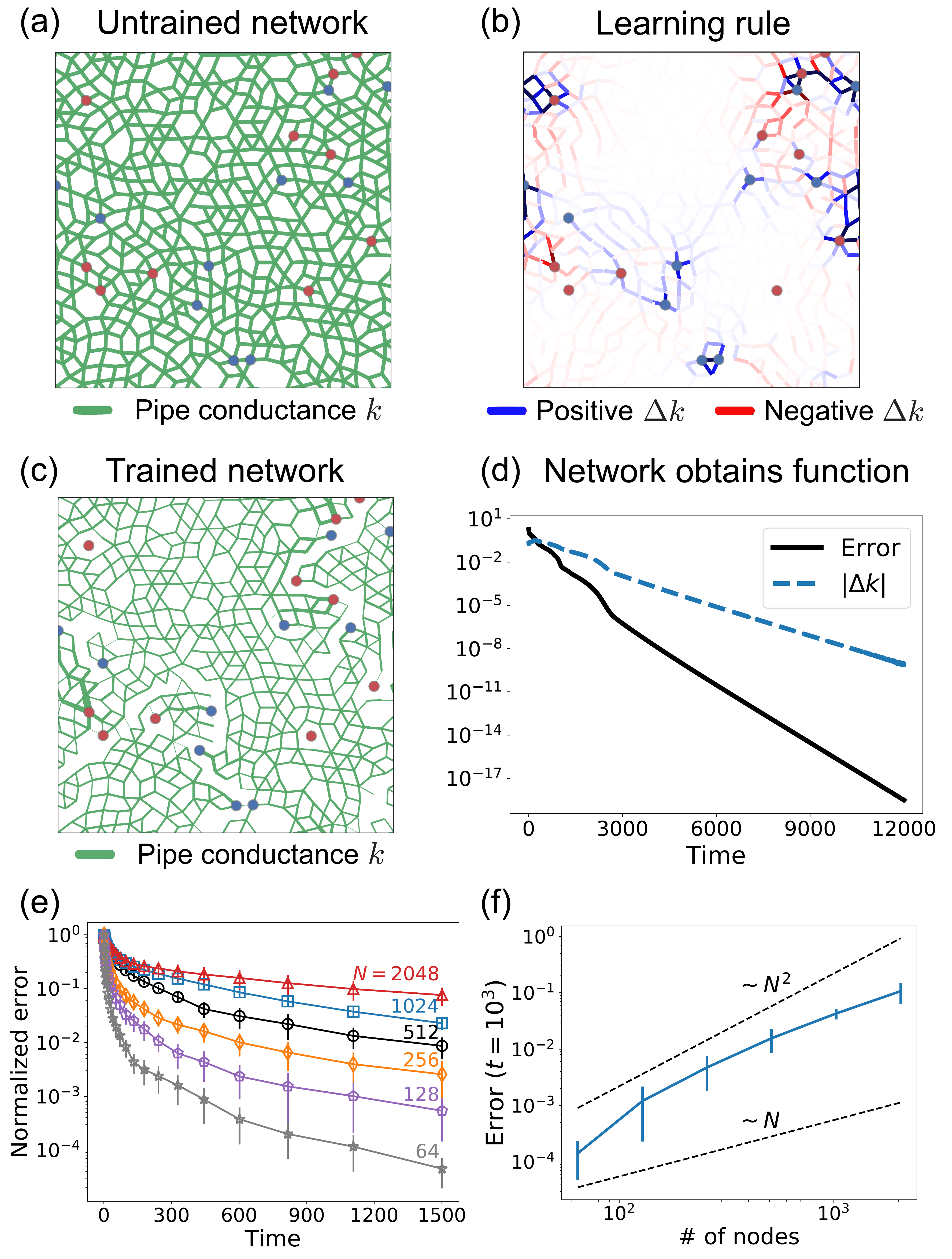}
\caption{Training flow networks with coupled learning. a) An untrained disordered flow network with uniform conductance at all pipes $k_i=1$, as indicated by uniform thicknesses of the green edges. The ten red and blue nodes correspond to the source and target nodes with dot sizes indicating the magnitudes of the source pressures $\{P_S\}$ and {\emph{desired}} target pressures $\{P_T\}$. b) In each step, conductance values are modified using Eq.~\ref{eq:FlowNetworkRule}, according to the difference in flow between the free and clamped states. This process is applied iteratively. c) After training, the network conductance values, indicated by the thicknesses of the green edges, are considerably changed compared to the initial network shown in (a). d) During training of a network ($N=512$ nodes), the pressure values of the target nodes approach the desired values, as indicated by the exponentially shrinking error (black solid line). The desired target values $\{P_T\}$ are reached when the error is small; the modification of the conductance in each time step, $\Delta k$ (blue dashed line), vanishes exponentially as well. e) We train multiple networks of different sizes $N=64-2048$, and find that all can be trained successfully with coupled learning. Error bars indicate the variation with initial network and choice of sources and targets. In all cases, errors decay exponentially, yet larger networks converge slower. f) Picking a certain time $t=10^3$, we find that the error scales up with system size as a soft power between $1$ and $2$.
\label{fig:TrainFlow}}
\end{figure}

We test the coupled learning rule, Eq.~\ref{eq:FlowNetworkRule}, on disordered flow networks of various sizes ($N=64,128,256,512,1024,2048$). Each of these networks is initialized with pipes of uniform conductance $k_j=1$ in Fig.~\ref{fig:TrainFlow}(a). In each instance, we pick $M_S=10$ nodes randomly to be the source nodes and apply the source pressures $P_S$, indicated by the sizes of the red dots. These source pressures are themselves chosen randomly from a Gaussian distribution $\mathcal{N}(0,1)$. We also randomly choose $M_T=10$ different target nodes, and desired pressures at target nodes, $\{P_T\}$, randomly from a Gaussian distribution scaled by the source pressures, $\mathcal{N}(0,0.2 \sum P_S)$; their values are indicated by the sizes of the blue dots in Fig.~\ref{fig:TrainFlow}(a). 
The free state pressures $p^F$, given applied source pressures $P_S$, are first computed by minimizing the dissipated power. To compare the output $p_T^{F}$ of the network to the desired response $P_T$, we use the standard error function $C=\frac{1}{2} \sum_T [p_T^{F}-P_T]^2$.  Untrained networks have a large error, and training should gradually decrease it. The network achieves the function perfectly when the error vanishes.

We now consider the clamped state by nudging the target nodes towards their desired values, setting their pressures to $\{p_T^C=p_T^{F}+\eta [P_T- p_T^{F}]\}$, with $\eta=10^{-3}$. We minimize power dissipation with respect to the hidden nodes, $p_H$, and then update the conductance values (learning degrees of freedom) according to Eq.~\ref{eq:FlowNetworkRule} with a learning rate of $\alpha=5\cdot 10^{-4}$. Fig.~\ref{fig:TrainFlow}(b) shows the change of conductance of each edge at the first iteration of learning, with blue (red) signifying positive (negative) conductance changes. This process constitutes one step of the training process; at the end of each step, we compute the error function $C$ (Fig.~\ref{fig:TrainFlow}(d). The difference between the obtained targets and the desired ones decreases exponentially by many orders of magnitude during the training process, reaching machine precision. This result demonstrates the success of the coupled learning approach. We see that the magnitude of the change in the conductance vector, $\vert \Delta k \vert =\sqrt{\sum_j \Delta k_j^2}$, calculated for each step of the training process, also decreases exponentially during training (blue dashed line in Fig.~\ref{fig:TrainFlow}(d)). This result shows that the learning process is adaptive--it slows down as it approaches good solutions. The final trained network is displayed in Fig.~\ref{fig:TrainFlow}(c), with edge thicknesses indicating conductance. The pipes of the trained network have changed considerably compared to the initial one shown in Fig.~\ref{fig:TrainFlow}(a), with some pipes effectively removed (with conductances near zero).

The results of applying the training protocol to networks of different sizes, for different initial networks and choices of the source and target nodes and their pressure values, are shown in Fig.~\ref{fig:TrainFlow}(ef), where errors are rescaled by the initial error for each network. Our learning algorithm is generally able to train the networks to exhibit the desired responses, successfully decreasing the initial error by orders of magnitude. 
We find that networks of different size converge at on good solutions at different rates, with the error at a particular chosen time $t=10^3$ scales roughly as a power law $C(N,t)\sim N^q$ (with power $q$ in the range $1-2$). We note that networks of different sizes may not be equivalent, as training may depend on idiosyncratic details, such as particular distances between sources and targets, or other geometrical features. We leave detailed exploration of the effects of network size and source-target geometry to future study.

It is noteworthy that flow networks are linear, so that the mapping between the sources to targets are always linear $p_T = A(k) P_S$ ($A$ is a $M_S\times M_T$ matrix that depends on the conductance values). Networks with contain hundreds of edges have many more conductance values than components of $A$ so that there are far more degrees of freedom than constraints. While this argument suggests our flow networks are over-parameterized and should always succeed in learning, we stress that not all linear transformations are possible; pressure values everywhere in the network are weighted averages of their neighbors (due to Kirchhoff's law). More importantly, the linear transformation is limited because all conductance values must be non-negative (see appendix D). As a result, low networks cannot implement any desired linear mapping between the inputs and outputs, and non-zero errors are expected for certain tasks after training. It was previously shown that the likelihood of flow networks to successfully learn a set of tasks depends on network size~\cite{rocks2019limits}, even when trained with gradient descent. Therefore, we expect that the training larger networks for a given task is more likely to succeed due to over-parameterization, at the expense of slower convergence rates.

Furthermore, while computational neural networks are often initialized with random (e.g. Gaussian distributed) weights to compensate for their symmetries~\cite{thimm1997high,yam2000weight}, we find that our disordered networks can be trained successfully with uniform $k_j=1$ initialization. Indeed, our tests with initial weights of the form $k_j=\mathcal{N}(1,\sigma^2)$, with $0 < \sigma \leq 0.5$, have yielded qualitatively similar results to those shown in Fig.~\ref{fig:TrainFlow}b, with respect to both training time and accuracy.


\added{Recently, a similar set of ideas~\cite{kendall2020training, laborieux2020scaling} based on equilibrium propagation was independently proposed to train electric resistor networks. This approach is very similar to our framework. The difference is that Refs.~\cite{kendall2020training, laborieux2020scaling} used a nudged state defined by injecting currents to target nodes, instead of by clamping voltages as in our approach. Their method converges to gradient descent in the limit $\eta\rightarrow 0$. While our method does not, we find that coupled learning and gradient descent give very similar results in our trained networks (See Section IV). Our focus in this paper is also somewhat different from theirs; we showcase coupled learning in general physical networks in which the system naturally tends to minimize a physical cost function with respect to the physical degrees of freedom, and emphasize considerations regarding the implementation of coupled learning in experimental physical networks.}


\subsection{Coupled Learning in general physical networks}

We now formulate coupled learning more generally so that it can be applied to general physical networks. Consider a physical network with nodes indexed by $\mu$ and edges indexed by $j$. As in the special case described earlier, $\{x_\mu\}$ are \textit{physical degrees of freedom} (e.g. node pressure for a flow network). Network interactions 
depend on the learning degrees of freedom on the network edges, $\{w_j\}$ (e.g. pipe conductance in flow networks). We restrict ourselves to the athermal case where, given some initial condition $x_\mu(t=0)$, the physical degrees of freedom evolve to minimize a physical cost function $E$ (e.g. the power dissipation for a flow network) to reach a local or global minimum $E(x_\mu;w_j)$ at $x_\mu$, where both $E$ and $\{x_\mu\}$ depend on $\{w_j\}$.

We define the \textit{free} state $x_\mu^{F}$ where the source variables $x_\mu$ alone are constrained to their values, and $x_T,x_H$ equilibrate (\textit{i.\,e.~} minimize $E(x_\mu;w_j)$). At the \textit{clamped} state $x_\mu^{C}$ both source and target variables $x_S,x_T$ are constrained, so only the hidden variables are equilibrated. An external trainer (supervisor) nudges the target nodes by a small amount $\eta$ towards the desired values $X_T$, 

\begin{equation}
 x_T^{C}=x_T^{F} + \eta [X_T - x_T^{F}] \\
 \label{eq:ClampeddefG}
\end{equation}
with $\eta \ll 1$. The general form of coupled learning is then:

\begin{equation}
\begin{aligned}
\dot{\bm{w}} = \alpha \eta^{-1} \partial_{\bm{w}} \{E^{F}(x_S,x_T^{F})
- E^{C}(x_S,x_T^{C})\},
\end{aligned}
  \label{eq:GeneralRule}
\end{equation}
where $\alpha$ is again the learning rate. We thus find a general learning rule similar to equilibrium propagation~\cite{scellier2017equilibrium}, with the understanding that only the direct derivatives ($\partial_{\bm{w}} E$) are performed, excluding the physical state derivatives $\partial_{\bm{w}} x^{F}, \partial_{\bm{w}} x^{C}$ which cancel out. Given any physical network with a known physical cost function $E(x_\mu;w_j)$, one can derive the relevant coupled learning rule directly from Eq.~\ref{eq:GeneralRule}. 
Coupled learning is performed iteratively until the desired function is achieved. Note that in physical networks, $E$ can generically be partitioned as a sum over edges $E=\sum_j E_j(x_\mu(j);w_j)$. Each term $E_j$ for edge $j$ depends only on the physical degrees of freedom attached to the two nodes $\mu(j)$ that connect to edge $j$. The learning rule of Eq.~\ref{eq:GeneralRule} is thus guaranteed to be local, in contrast to many design schemes relying on the optimization of a global cost function. Moreover, the network is not required to encode any information about the desired response. This information is fed to the network by the trainer, which nudges the target physical degrees of freedom closer to the desired state at every iteration of the learning process. Note that while our discussion so far and Fig.~\ref{fig:Schematic} are restricted to a physical network with binary edges (i.e. binary physical interactions), coupled learning does not assume binary interactions and is valid for arbitrary $n$-body potentials. For more information on the general coupled learning rule, see Appendix B.


\subsection{Elastic networks}

To demonstrate the generality of our coupled learning framework, we apply it to another physical system, central-force spring networks. 
Here we have a set of $N$ nodes embedded in $d$-dimensional space, located at positions $\{x_\mu\}$. 
The nodes are connected to their neighbors in the network by linear springs, each having a spring constant $k_j$ and equilibrium length $\ell_j$. 
The energy of a spring, labeled $j$, depends on the strain of that spring $E_j=\frac{1}{2}k_j(r_{j}-\ell_j)^2$, where $r_{j}$ is the Euclidean distance between the two nodes connected by the spring. The physical cost function in this case is the total elastic energy of the network, given by $E=\sum_{j\in springs}E_j$. 
We choose source and target springs, whose lengths are $r_S$ and $r_T$, and train the network so that an application of source edge strains $\{R_S\}$ gives rise to desired target edge strains $\{R_T\}$.

In contrast to flow networks, spring networks are non-linear in the physical variables $\{x_\mu\}$, due to the non-linearity in the Euclidean distance function. 
Moreover, while the spring constants $\{k_j\}$ are formally equivalent to conductances in flow networks, the equilibrium lengths $\{\ell_j\}$ have no direct analog in flow networks. These extra variables are additional learning degrees of freedom that we can adjust in addition to the spring constants. 

\begin{figure}
\includegraphics[width=1.0\linewidth]{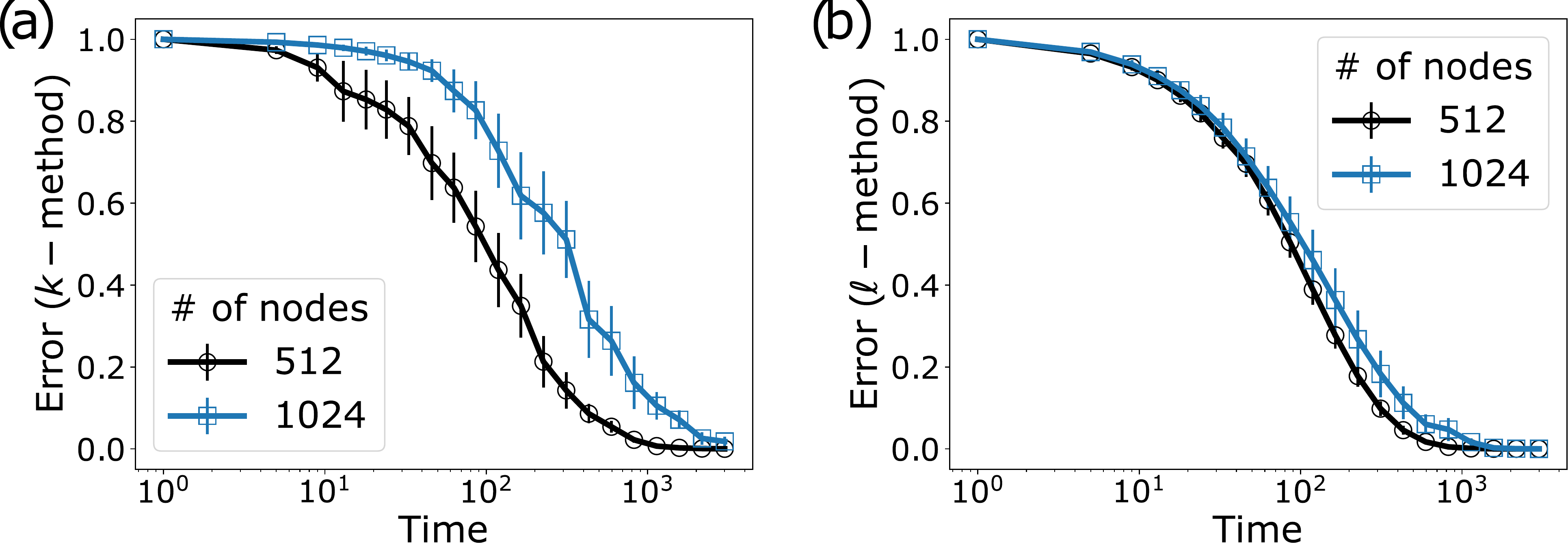}
\caption{Training spring networks by modifying (a) spring constants or (b) rest lengths. The error in the target edge lengths is shown as a function of the number of training iterations for $N=512$ (black circles) and $N=1024$ (blue squares) networks. Training by coupled learning is successful for either the stiffness and rest length degrees of freedom.
\label{fig:ElasticNets}}
\end{figure}

The free and clamped states of the spring network are applied similarly to the previous example, with the exception that we define the source and target boundary conditions on the edges of the network rather than the nodes, demonstrating that the coupled learning framework can be applied in either case. Next, we apply the coupled learning rule~\ref{eq:GeneralRule} to obtain two separate update rules, one for the spring constants $k_j$, the other for the rest lengths $\ell_j$:

\begin{equation}
\begin{aligned}
\dot{k_j} =&\frac{\alpha}{\eta} \frac{\partial}{\partial k_j}\{E^{F}
-  E^{C}\} =\\ =&\frac{\alpha}{2\eta} \{[r_j^{F}-\ell_j]^2
-  [r_j^{C}-\ell_j]^2\}, \\
\dot{\ell_j} =& \frac{\alpha}{\eta} \frac{\partial}{\partial \ell_j}\{E^{F}
-  E^{C}\}\\
=& \frac{\alpha }{\eta} k_j \{[r_j^{F}-\ell_j]
- [r_j^{C}-\ell_j]\}.
\end{aligned}
  \label{eq:ElasticNetworkRule}
\end{equation}

Learning in elastic networks can be accomplished through modification of the spring constants, rest lengths, or a combination of both. As before, Eq.~\ref{eq:ElasticNetworkRule} gives local learning rules, where each spring only changes due to the strain on that particular spring. To test these learning rules, we train elastic networks with $N=512$ and $N=1024$ nodes with multiple choices of $10$ random source strains and $3$ random target strains. Elastic network calculations were performed using a specialized bench-marked set of computational tools, used earlier for research on tuned networks~\cite{goodrich2015principle,rocks2017designing}. The success of the training is again assessed using the error $C=\frac{1}{2} \sum_T [r_T^{F}-R_T]^2$, measuring deviation from the desired target lengths. We find that regardless of whether the network learns by modifying its spring constants (Fig.~\ref{fig:ElasticNets}(a)) or rest lengths (Fig.~\ref{fig:ElasticNets}(b)), the networks are consistently successful in achieving the desired target strain, reducing the error $C$ by orders of magnitude. Larger networks take slightly longer to learn, but achieve similar success. We find that the rest-length based learning rule gives somewhat more consistent learning results for different networks and source-target choices, as evidenced by smaller error bars in Fig.~\ref{fig:ElasticNets}b. \added{Recently, a rule similar to that of Eq.~\ref{eq:ElasticNetworkRule} was used to prune edges in elastic networks to obtain desired allosteric responses~\cite{pashine2021local}. It is notable that~\cite{pashine2021local} uses a `symmetrized' version of the learning rule, where the free state is replaced with a negative clamped state, where the output nodes are held farther away from their respective desired value.}

\subsection{Supervised classification}

\begin{figure}
\includegraphics[width=1.0\linewidth]{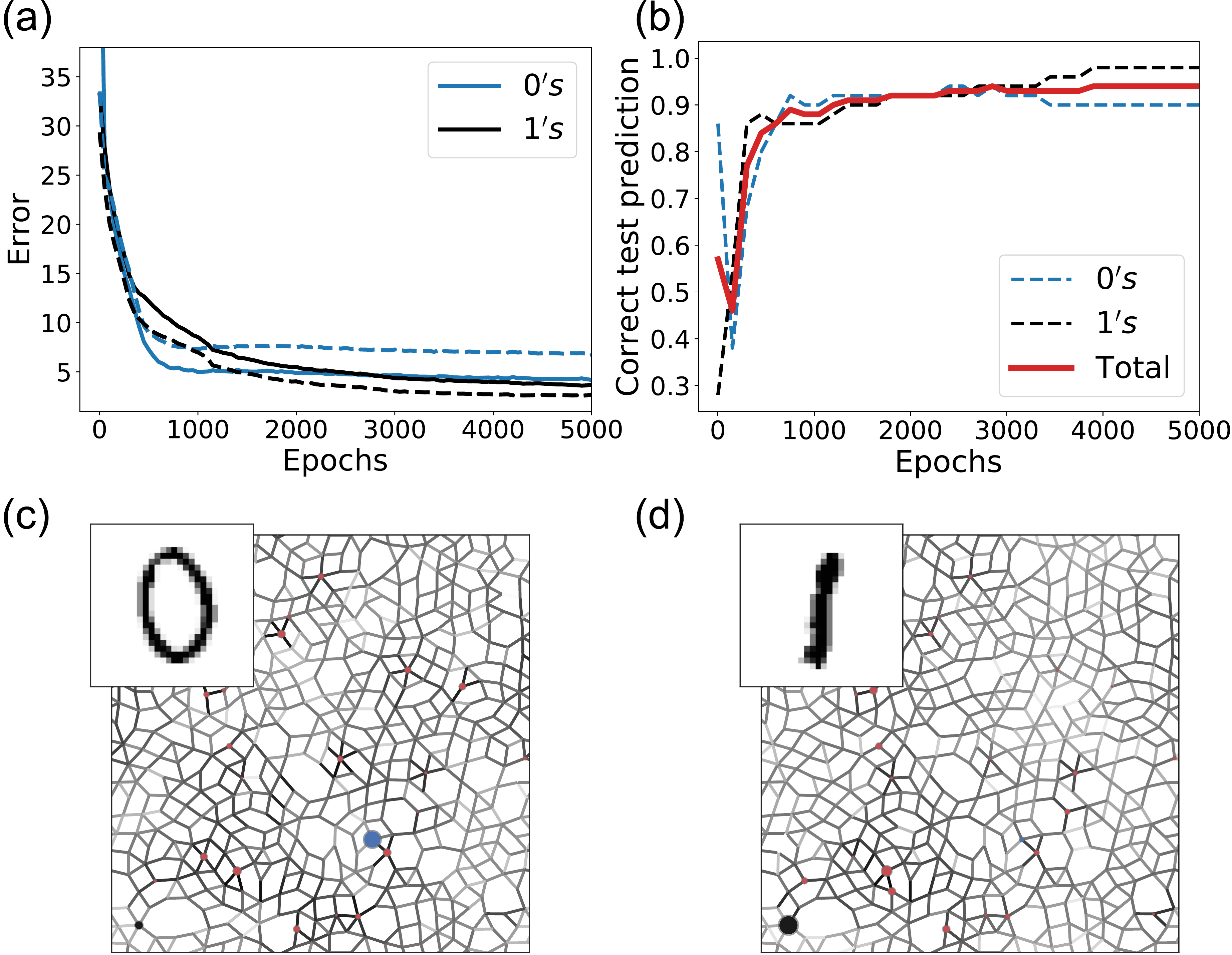}
\caption{Classification of MNIST digits (0's \& 1's). a) The classification error (value of cost function) (Eq.~\ref{eq:CostFunctionMain} for the digits 0 (blue) and 1 (black) vs. the number epochs, where each epoch consists of 100 iterations of the learning rule with a different randomly selected training image presented at each iteration. Training error is indicated by solid lines and test error by dashed lines. b) Prediction accuracy on test images (fraction of correct predictions made by the network for digits 0 and 1) vs. number of epochs. Total test accuracy shown in red. (c-d) Response of the network when presented with the image at the top-left. The power dissipated at each edge is indicated by the intensity of the edge color. while the pressures of target nodes are labeled with blue (0) and black (1) circles. When a 0 image is shown, the 0 target node has high pressure (blue circle) and the 1 target node has low pressure; when a 1 image is shown, the 1 target node has high pressure (black circle) and the 0 target node has low pressure.
\label{fig:MNIST}}
\end{figure}

In Secs. II.A and C we trained networks to exhibit functions inspired by allostery in biology, obtaining a desired map between one set of sources (inputs) and one set of targets (outputs). Here we train flow networks to display a function inspired by computer science, namely the ability to classify images. In classification, a network is trained to map between multiple sets of inputs and outputs. In order to use our coupled learning protocol for simultaneous encoding of multiple input-output maps, we slightly modify the training process in the spirit of stochastic gradient descent~\cite{bottou2003stochastic}. Each training image provides an input and the correct answer for that image is the desired output. Training examples are sampled uniformly at random, and the coupled learning rule of Eq.~\ref{eq:FlowNetworkRule} is applied accordingly.

A simple example of a classification problem, often used to benchmark supervised machine learning algorithms, is to distinguish between labeled images of handwritten digits (MNIST~\cite{lecun2010mnist}). Each image corresponds to a particular input vector (e.g. pixel intensity values), while the desired output is an indicator of the digit displayed. Typically an algorithm is trained on a set of example images (training set), while the goal is to optimize the classification performance on unseen images (test set).

Here, we train flow networks to distinguish between images of two digits ($0$ and $1$). We pick $50$ images of each digit (class) from the MNIST database to serve as a training set, and an additional $50$ images of each digit as a test set. Instead of using the images themselves, we first carry out a Principal Component Analysis (PCA) of all MNIST images, and train the network using the top 25 principal components of the training set images. The inputs (source pressures) are given by these principal components of the training images for a randomly selected set of 25 source nodes. We additionally pick $2$ random output (target) nodes, one for each digit. The desired output for a training example corresponding to a `$0$' is for the first output node, the `$0$' target node to have high pressure ($p_{`0'}=1$), and for the second output node, the `$1$' node, to have no pressure ($p_{`1'}=0$). The target pressures are reversed when an example corresponding to the digit `$1$' is chosen so that the `$1$' node has $p_{`1'}=1$ and the `$0$' node has $p_{`0'}=0$. At each iteration of the learning rule, the network is presented with a single training image, sampled uniformly at random from the training set, and is nudged towards the desired output for that image. A training epoch is defined as the time required for the network to be presented with $100$ training examples.

The error between the desired and observed behavior is shown in Fig.~\ref{fig:MNIST}(a) for each of the digits. The solid lines represent errors for the training set, while the dashed lines show errors for the test set. We see that the network not only reduces the training error, but also successfully generalizes, reducing the test errors as well. While the pressure at the output nodes is continuous, we want the network to make a discrete choice about the label of the shown image. We say that the predicted label is given by the target node with the larger pressure. With this definition, we see that the classification accuracy of the network starts at about $50\%$ and improves dramatically during training (Fig.~\ref{fig:MNIST}b). After training, the overall training accuracy reaches $98\%$, while the test accuracy is $95\%$. This learning performance is almost as good as that of simple machine learning algorithms such as logistic regression, which for the same data yields a training accuracy of 100\% and a test accuracy of 98\%. Note that we did not tune hyper parameters to achieve the listed accuracies; such tuning might well improve the performance of our algorithm. In Fig.~\ref{fig:MNIST}(c-d), the network response to two select input images is shown. When the source nodes are constrained with pressure values corresponding to an image of a `$0$', the `$0$' target node has high pressure (blue) while the `$1$' target node has very low pressure. The opposite response occurs when an image of a `$1$' is applied as input.

As discussed above, we used a training approach in the spirit of stochastic gradient descent, training the network with one training example at a time. We note that it may be experimentally possible to train the network using batch gradients, presenting many training examples in quick succession, so that the learning rule averages the conductance update over these examples. A similar effect can be achieved by maintaining the rate of example presentation, yet slowing down the learning rate $\alpha$.

\section{Experimental considerations}

In the previous section we introduced the coupled learning rules that respect the physical requirements of locality and the fact that the desired response cannot be encoded in the network \textit{a priori}. We have demonstrated computationally that flow and mechanical networks successfully learn desired behaviors using such learning rules. For networks to learn autonomously in a physical setting, the network must be able to implement the coupled learning rules as the physical dynamics of their learning degrees of freedom. Furthermore, an external supervisor needs to be able to clamp the target nodes as needed. For fluid flow networks, coupled learning requires the coupling of pipe conductance values to the flow inside the pipe. For example, constriction of pipes can be achieved by entrainment of sediment in the flow, and its deposition inside the pipes~\cite{soo1972deposition,uijttewaal1996particle}. These dynamics could support a one-sided learning rule as described in Sec III.A. In elastic networks, either the stiffnesses or rest lengths of bonds need to change for learning to be possible. Several classes of materials are known to adapt their stiffness in response to applied strain, e.g. ethylene vinyl acetate (EVA)~\cite{jin2010uv} and thermoplastic polyurethane~\cite{boubakri2010impact}. The former has been used to train materials for auxetic behavior using directed aging~\cite{pashine2019directed}.

In the following we describe some potential experimental difficulties in implementing coupled learning in physical networks networks and how they may be overcome. We first discuss an approximation to the coupled learning rules, allowing for a simpler implementation in flow networks (Sec. III.A). Then, we describe the effects of plausible experimental issues, such as noise (Sec. III.B), drifting learning degrees of freedom (Sec. III.C) and the effect of network size on the physical relaxation assumed in coupled learning (Sec. III.D).

\subsection{Approximate learning rules}

Consider the flow networks described earlier. To implement the learning rule of Eq.~\ref{eq:FlowNetworkRule}, one requires pipes whose conductance can vary in response to the dissipated flow power through them. There are two major issues in implementing this learning rule in a physical flow network. First, the learning rule of Eq.~\ref{eq:FlowNetworkRule} for the conductance on edge $j$ depends not simply on the power through edge $j$ but on the difference between the power when the free boundary condition and the power when the clamped boundary condition are applied. This is difficult to handle experimentally because it is impossible to apply both sets of boundary conditions simultaneously to extract the difference. One could try to get around this by alternating between the free and clamped boundary conditions during the process, but then one runs into the difficulty that the sign of conductance change is opposite for the two types of boundary conditions. In other words, the same power through edge $j$ in the free and clamped states would need to induce the opposite change in conductance. Alternating the sign of the change in conductance along with the boundary conditions poses considerable experimental difficulty. The second hurdle is that Eq.~\ref{eq:FlowNetworkRule} requires that pipes must be able to either increase or decrease their conductance. Pipes whose conductance can change in only one direction (e.g. decreasing $k$ by constriction) are presumably  to implement experimentally.

To circumvent these difficulties, we seek an appropriate approximation of Eq.~\ref{eq:FlowNetworkRule} (see Appendix C for more details) that is still effective as a learning rule. We first define a new hybrid state of the system, named the $\delta$-state, in which the power dissipation is minimized with respect to the hidden node pressures $p_H$ with the constraints $p_S=0$, $p_T=-\eta(P_T-p_T^{F})$.

Note that this state corresponds to constraining the source and target nodes to the desired pressure differences between the free and clamped states according to the original approach. Now, we may expand the clamped state power in series around the free state, to obtain a new expression for the learning rule in terms of the $\delta$-state and free state:

\begin{equation}
\begin{aligned}
\dot{k}_j \approx 2\alpha\eta^{-1} \Delta p_j^{F} \Delta p_j^{\delta}. \nonumber
\end{aligned}
  \label{eq:DeltaRule0}
\end{equation}

Ideally only the $\delta$ state would be involved in the learning rule and not the free state. 
\added{Let us simplify this rule by only accounting for the sign of the free pressure drop $\textrm{sgn}(\Delta p_j^{F})$, returning $\pm 1$ depending on the sign of flow in the free state $\dot{k}_j \approx 2\alpha\eta^{-1} \textrm{sgn}(\Delta p_j^{F}) \Delta p_j^{\delta}$.}

The resulting learning rule, while only depending on the pressures at the $\delta$-state, can still induce either positive or negative changes in $k$. To avoid this, we may impose a step function, that only allows changes in one direction (e.g. only decreasing $k$):


\added{\begin{equation}
\begin{aligned} 
\dot{k}_j = -2\alpha\eta^{-1} \Theta(-\Delta p_j^{F}\Delta p_j^{\delta}) \vert \Delta p_j^{\delta}\vert .
\end{aligned}
  \label{eq:DeltaRule}
\end{equation}}

\begin{figure}
\includegraphics[width=1.0\linewidth]{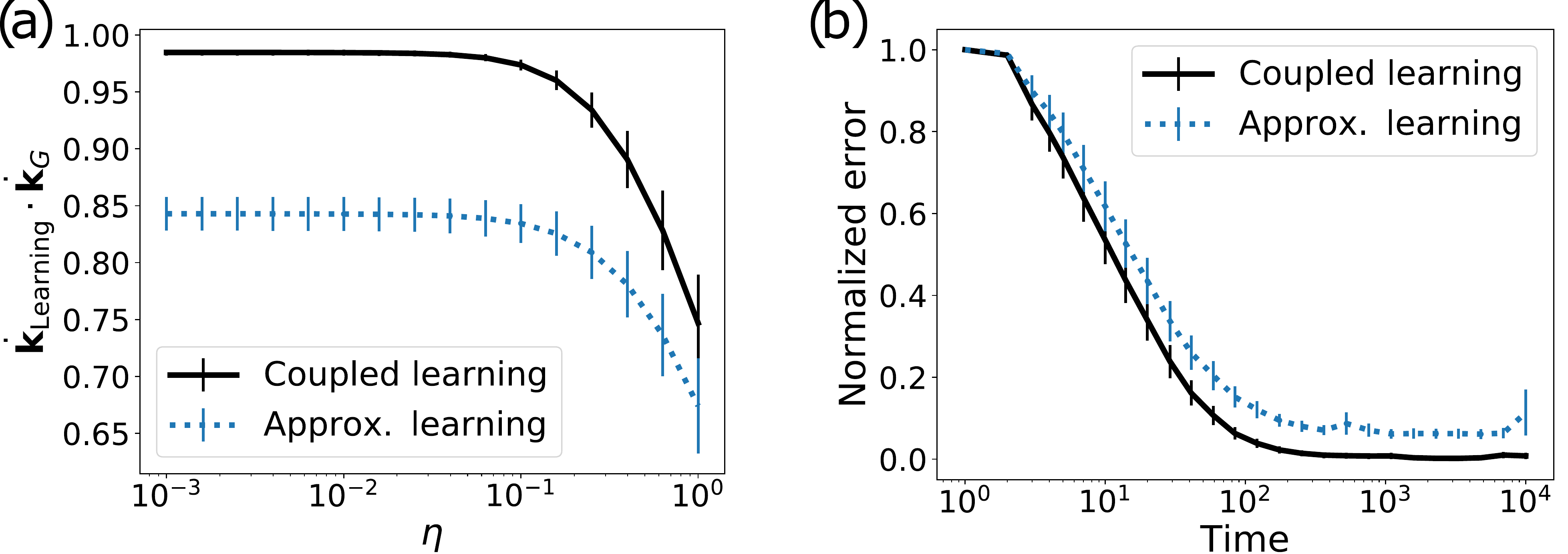}
\caption{Training a flow network ($512$ nodes, $10$ random sources and $3$ random targets) with an approximate, experimentally viable learning rule. a) The full coupled learning rule (Eq.~\ref{eq:FlowNetworkRule}) and the approximate learning rule (Eq.~\ref{eq:DeltaRule}) are compared to the gradient of the cost function for different values of $\eta$ (see Fig.~\ref{fig:LearnVTune}a). While the full learning rule is superior, there is still a significant correlation between optimal tuning and the approximate rule. b) The network may be trained using both the full and approximate rule, at similar time scales. The approximate learning rule, being much more restricted, often saturates at a higher error level compared to the full rule (averaged over $17$ random source-target sets).
\label{fig:ExpRule}}
\end{figure}

The intuition behind this learning rule is relatively simple: if for a particular pipe $\textrm{sgn}(\Delta p_j^{F}) \Delta p_j^{\delta}<0$, then too much flow is going through the pipe in the free state. Thus, one could improve the functional result by decreasing the conductance of that pipe.

This learning rule may be simpler to implement physically, but is it a sufficiently close approximation of the original rule? In Fig.~\ref{fig:ExpRule} we train a network ($512$ nodes, with $10$ random sources and $3$ random targets) using the learning rules of Eqs.~\ref{eq:FlowNetworkRule} and~\ref{eq:DeltaRule}. In Fig.~\ref{fig:ExpRule}a, the learning rules are compared to the tuning algorithm (Eq.~\ref{eq:TuningVSLearning}) by taking the dot product of these modification (similar to Fig.~\ref{fig:LearnVTune}a). While the approximate learning rule has a lower dot product with the optimal local tuning for all values of $\eta$ compared to the full learning rule, the correlations are still very significant. We thus expect the approximate learning rule to train the network successfully. Indeed, we find that when the approximate learning rule is applied iteratively, the learning process is successful, yielding a decrease of the error $C$ of at least an order of magnitude  (Fig.~\ref{fig:ExpRule}(b)). While the approximate training typically plateaus at a higher error level compared to the full coupled learning rule, the approximate learning rule can successfully train flow networks to exhibit random source-target responses.

\subsection{\added{Noise in physical or learning degrees of freedom}}

So far, we have assumed that coupled learning modifies the learning degrees of freedom given precise knowledge of the free and clamped states (e.g. $\mathcal{P}^F,\mathcal{P}^C$ in Eq.~\ref{eq:FlowNetworkRule}). However, in any physical implementation of a network there will be errors in the measurements of these states. Suppose that measurement of the dissipated power at every edge is subject to additive white noise of magnitude $\epsilon$, noted $e_j\sim  \mathcal{N}(0,\epsilon^2)$. The learning rule would then contain an extra term due to the noise

\begin{equation}
\begin{aligned}
\dot{k}_j \approx \frac{1}{2}\alpha \eta^{-1} \{[\Delta p^{F}_j]^2 - [\Delta p^{C}_j]^2 + e_j\}.
\end{aligned}
  \label{eq:FlowNetworkRuleNoise}
\end{equation}

It is clear from Eq.~\ref{eq:FlowNetworkRuleNoise} that if the error magnitude $\epsilon$ is larger than the typical difference between the free and clamped terms, the update to the conductance values would be random and learning will fail. As we show in Appendix B, the difference between the free and clamped term scales with the nudge amplitude $[\Delta p^{F}]^2 - [\Delta p^{C}]^2 \sim \eta^2$. Put differently, when the nudge is small ($\eta\rightarrow 0$), the free and clamped states are nearly identical, so that their difference is dominated by noise. This raises the possibility that increasing the nudge amplitude $\eta$ used, i.e. nudging the clamped state closer to the desired state, will increase the relative importance of the learning signal compared to the noise, allowing the network to learn despite the noise.

\begin{figure}
\includegraphics[width=1.0\linewidth]{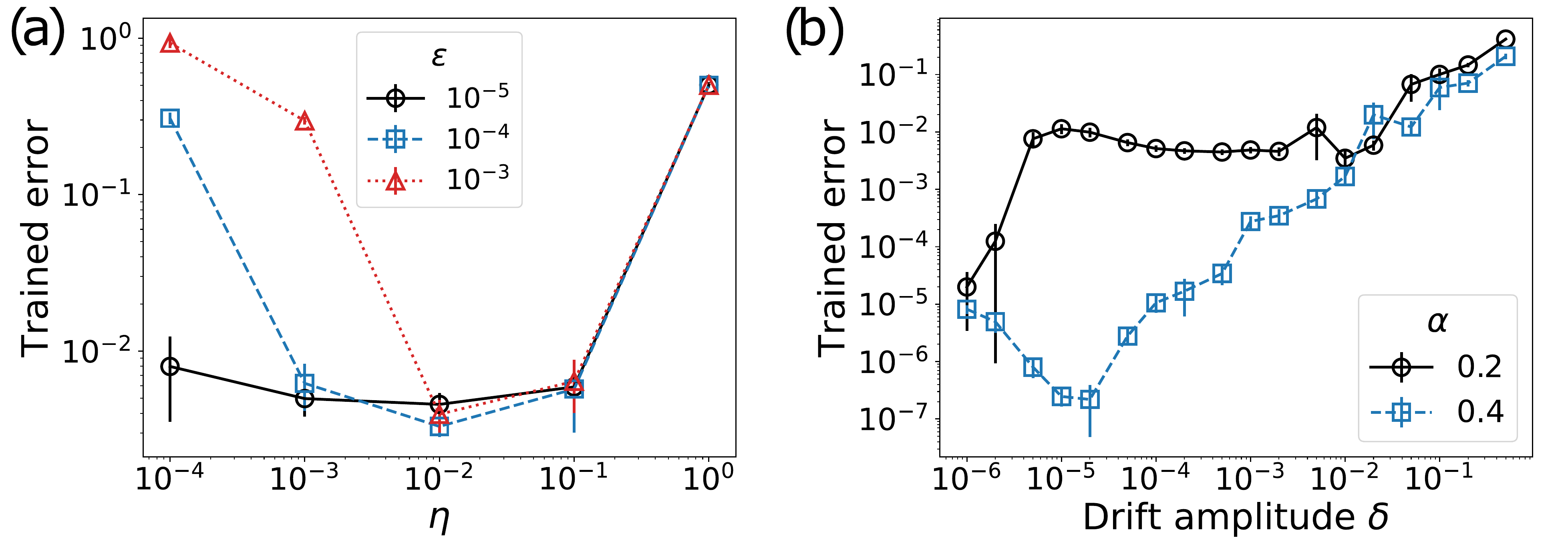}
\caption{Learning with noisy measurements and diffusing learning degrees of freedom. a) When noise is present in the free or clamped state measurements, small nudges impede learning. Training results can be improved with larger $\eta<1$ nudges. b) Drifting in the learning degrees of freedom impedes learning, but its effects can be mitigated by increasing the learning rate $\alpha$.
\label{fig:ExpLim}}
\end{figure}

To test this idea, we train a flow network of $N=512$ nodes on a random task with $10$ source and $10$ target nodes (the same training problem shown in Fig.~\ref{fig:TrainFlow}a-d). While training, we inject white noise of given magnitude $\epsilon$ according to Eq.~\ref{eq:FlowNetworkRuleNoise}. As suggested, we find that these `measurement' errors primarily spoil learning at low nudges values, so that increasing $\eta$ might have a beneficial effect on learning (Fig.~\ref{fig:ExpLim}a). We conclude that experimental settings with finite measurement errors induce a trade-off in the nudge parameter $\eta$. Nudges should be small so that learning rule gives a good estimate of the desired gradients, yet they can't be too small, or the learning will be dominated by noise.

Another source of error in an experimental realization of learning networks may be unwanted fluctuations or drift in the learning parameters. For example, it is known that natural neural network undergo synaptic fluctuations, even in adults and in the absence of synaptic activity~\cite{holtmaat2009experience,yasumatsu2008principles,stettler2006axons}. In synthetic contexts, engineering materials with controllable parameters in a wide range, e.g. variable stiffness in shape memory alloys~\cite{ratna2008recent}, can lead to an increased volatility in these controlled parameters~\cite{kuder2013variable}. 

Drifting in the learning degrees of freedom during the training process is expected to impede learning. To study its effects, we assume that the learning degrees of freedom fluctuate with some magnitude $\delta$ for all edges. This gives rise to additive white noise $d_j\sim \mathcal{N}(0,\delta ^2)$ in the update rule

\begin{equation}
\begin{aligned}
\dot{k}_j \approx \frac{1}{2}\alpha \eta^{-1} \{[\Delta p^{F}_j]^2 - [\Delta p^{C}_j]^2\} + d_j.
\end{aligned}
  \label{eq:FlowNetworkRuleDiffusion}
\end{equation}

In contrast to the noise in the physical degrees of freedom discussed earlier, such diffusion cannot be controlled by increasing the nudge amplitude. However, one can increase the learning rate $\alpha$ to hasten the learning dynamics compared to diffusion (keeping in mind that to maintain convergence, learning cannot be made too fast~\cite{mehta2019high}).

Once more, we train a flow network ($N=512$ nodes) on a task with $10$ random sources and $10$ random targets (same task as in Fig.~\ref{fig:TrainFlow}a-d). The network is trained with different diffusion magnitudes $\delta$ and two different learning rates $\alpha=0.2, 0.4$. We find that given constant diffusion rates, training is more successful at a higher learning rate, as expected (Fig.~\ref{fig:ExpLim}b). These results show that unavoidable experimental issues such as noise in the physical and learning degrees of freedom can be mitigated by controlling learning hyper-parameters such as the nudge amplitude $\eta$ and the learning rate $\alpha$.

\subsection{Physical relaxation}

In this work, we have taken the quasi-static limit of coupled learning, so that before computing either the free or the clamped state, the network first reaches its stable equilibrium (or steady state). In other words, we explored the limit where learning is much slower than the physical dynamics~\cite{ernoult2019updates}. Physical networks however take finite time to relax to their equilibrium, which does not scale with the amplitude of perturbation $\eta$ (in the linear response regime). Therefore, using small nudges in general does not help with accelerating the physical relaxation of the network towards its equilibrium states.

Moreover, the nudging between the free and clamped states induces a local perturbation, whose information content must propagate to the entire network before it can equilibrate. In both flow networks and elastic spring networks, the time for this information propagation is set by the speed of sound. In flow networks, the speed of sound depends on the compressibility of the fluid. Similarly, elastic spring networks have a finite speed of sound depending on their node masses, edge spring constants and network structure. \added{This information propagation time implies a linear lower bound on the scaling of the relaxation time with system size. Depending on the physics of the network and its architecture, relaxation time might scale more slowly with size. For example, in large branched dissipative flow networks the response time to a perturbation scales quadratically~\cite{fancher2021tradeoffs}.}

Consider a flow or elastic network of linear dimension $L$, trained by waiting time $\tau$ for relaxation before updating the learning degrees of freedom in Eq.~\ref{eq:GeneralRule}. Due to the aforementioned considerations, training a larger network of linear dimension $L'$ would require waiting \added{$\tau' \approx \tau f(L'/L)$ (with $f$ a faster than linear function)} for similar relaxation, so that the overall training time should scales with the physical length of the network. Note that this physical time scaling is different from the scaling of learning convergence with system size discussed earlier. \added{At the limit of small enough learning rates $\alpha$, when the learning dynamics effectively approximate gradient descent on the cost function, we argue that rescaling the learning rate in Eq.~\ref{eq:GeneralRule} by $\alpha' = \alpha f(L'/L)$ would counteract this increase in training time. However, the learning rate cannot increase without bound for two reasons. First, at high learning rates, learning may not converge as the gradient step may overshoot the minimum of the cost function, as is often the case in optimization problems~\cite{mehta2019high}. Furthermore, increasing the learning rate compared to the physical relaxation rate would eventually break the quasi-static assumption in Eq.~\ref{eq:GeneralRule}.} We leave the effects of breaking the assumption of quasi-staticity to future study.


\section{\added{Comparison of coupled learning to other learning approaches}}


To emphasize the advantages of our local supervised learning approach, we compare it to global supervised learning, where we adjust the learning degrees of freedom to minimize a learning cost function (or error function). Such a rational tuning strategy was previously demonstrated to be highly successful~\cite{rocks2019limits} and is fundamental to supervised machine learning~\cite{mehta2019high}. Here we compare coupled learning and tuning in the contexts of flow and elastic networks.

One usually defines the learning cost function as the distance between the desired response $P_T$ and the actual target response $p_T^{F}$. A commonly chosen function is the $L_2$ norm (Euclidean distance)

\begin{equation}
\begin{aligned}
{\mathcal{C}}\equiv \frac{1}{2}[p_T^{F}-P_T]^2.
\label{eq:CostFunctionMain}
 \end{aligned}
\end{equation}

A straightforward way to minimize this function with respect to the learning degrees of freedom (pipe conductance values $\{k_j\}$) is to perform gradient descent. In each step of the process, the source pressures $\{P_S\}$ are applied, the physical cost function (total dissipated power) is minimized to obtain the pressures $\{p_T^{F}\}$, the gradient of the learning cost function is computed, and the conductances $\{k_j\}$ are changed according to

\begin{equation}
\begin{aligned}
&\partial_{\bm{k}}{\mathcal{C}} = [p_T^{F}-P_T] \cdot \partial_{\bm{k}}p_T^{F} \\
&\dot{\bm{k}}_G = -\alpha \partial_{\bm{k}}{\mathcal{C}}
\label{eq:CostGradientMain}
 \end{aligned}
\end{equation}

where $\dot{\bm{k}}_G$ denotes the change in the pipe conductance predicted by the tuning (global supervised learning) strategy. Note that such a process cannot drive autonomous adaptation in a physical network-- Eq.~\ref{eq:CostGradientMain} cannot be a physical learning rule--for two fundamental reasons. First, the update rule is not local. Generally, the target pressure $p_T^F$ depends on the conductance of every pipe in the network, so each component of $\partial_{\bm{k}}{\mathcal{C}}$ contains contributions from the entire network. The modification of each pipe conductance depends explicitly on the currents through all of the other pipes, no matter how distant they are. Second, the tuning cost function ${\mathcal{C}}$ depends explicitly on the desired response $P_T$. Thus, if the network computes the gradient, it must encode information about the desired response. A random disordered network is not expected to encode such information {\it a priori}. Together, these two properties of the tuning process imply that this approach requires both computation and the modification of pipe conductance values by an external designer -- it cannot qualify as a physical, autonomous learning process.

The second point above, that the tuning cost function depends explicitly on the desired behavior $P_T$, deserves further discussion. In coupled learning, the desired target values $P_T$ do not appear explicitly in the learning rule of Eq.~\ref{eq:FlowNetworkRule}, but do appear explicitly in the definition of the clamped state (Eq.~\ref{eq:Clampeddef1}). This is a subtle but crucial distinction between coupled learning and tuning. In coupled learning, a trainer (supervisor) imposes boundary conditions only on the target nodes. The physics of the network then propagates the effect of these boundary conditions to every edge in the network. Then, each edge independently adjusts its learning degrees of freedom taking into account only its own response to the applied boundary conditions. In other words, the boundary conditions, imposed by the trainer, depend on the desired response $\{P_T\}$ but the equations of motion of the pipe conductance values $\{k_j\}$ themselves do not require knowledge of $\{P_T\}$ once the boundary conditions are applied. The trainer needs to know the desired network behavior, but the network itself does not. In the tuning process, by contrast, the pipe conductance values evolve according to an equation of motion that depends explicitly on $\{P_T\}$. Thus, tuning a physical network requires external computation of $\partial _k \mathcal{C}$ and the subsequent modification of the learning degrees of freedom by an external agent.

The difference between local and global supervised learning has fueled longstanding debates on how biological networks learn, and their relation to computational tuning approaches such as machine learning algorithms~\cite{richards2019deep}. While natural neural systems are complicated and can perform certain computations, simple physical networks such as the ones studied here definitely cannot. In order for these networks to adapt autonomously to external inputs and learn desired functions from them without performing computations, we need a physically plausible learning paradigm such as the coupled learning framework presented here. 

\begin{figure}
\includegraphics[width=1.0\linewidth]{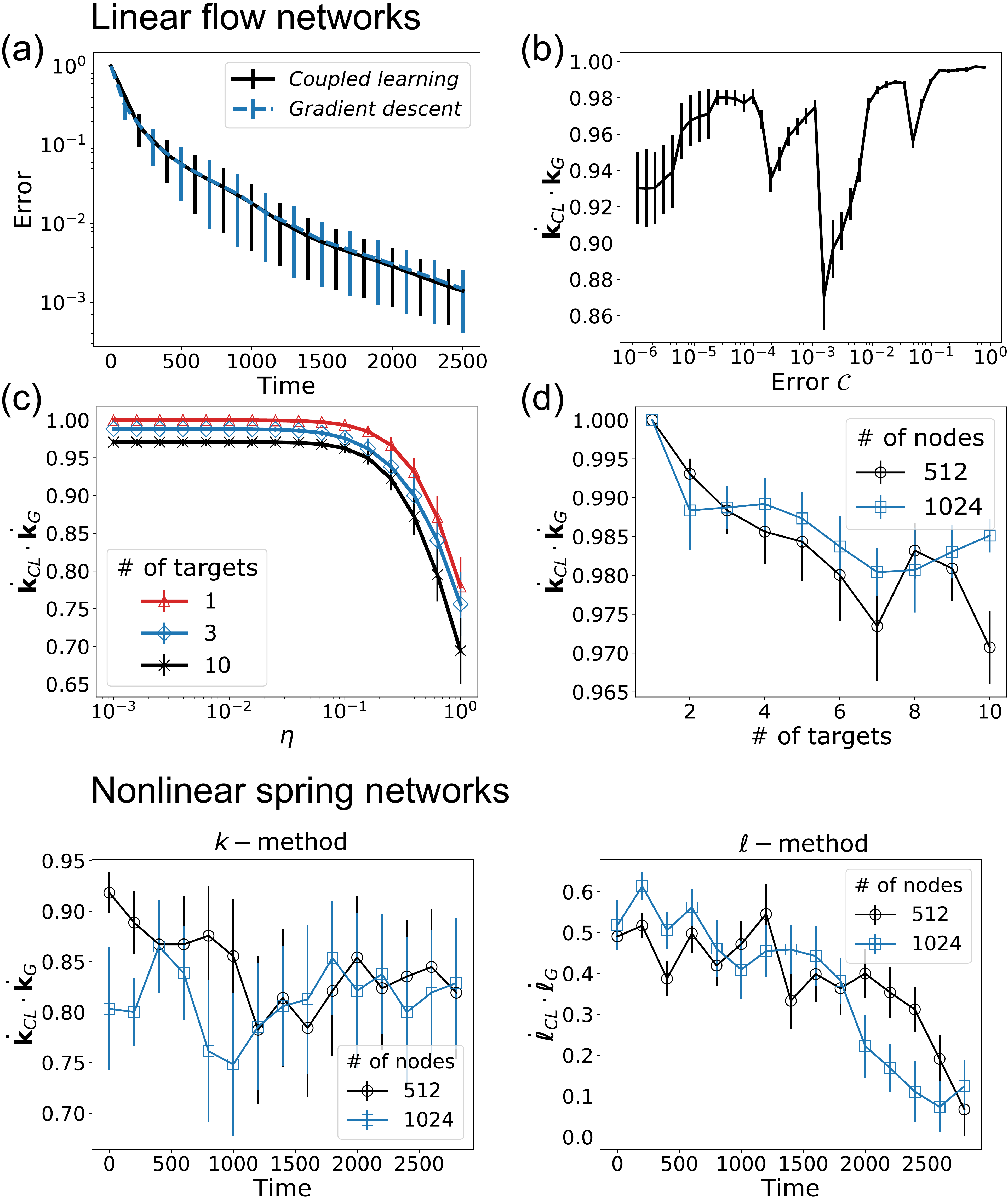}
\caption{Coupled learning (local supervised learning) compared to tuning (global supervised learning) in linear flow and nonlinear elastic networks. (a) Direct comparison between training traces in flow networks shows that coupled learning is as successful as gradient descent. (b) The dot product of the modification vector $\Delta\bm{k}$ for coupled learning and tuning (gradient descent) during training undergoes fluctuations, yet remains high. (c) At the beginning of training, we compute the dot product of the modification vector $\Delta\bm{k}$ for coupled learning and tuning as a function of the nudge variable $\eta$ for systems with 1 (red triangles), 3 (blue diamonds) and 10 (black crosses) target nodes. The two methods predict very similar results for $\eta\ll 1$. (d) For $\eta=10^{-3}$, coupled learning yields similar results to tuning for networks with $N=512$ (black circles) and $N=1024$ (blue squares), as a function of the number of target nodes. (e-f) Dot product between the normalized predictions of coupled learning and gradient descent in nonlinear elastic networks. We find that coupled learning and gradient descent are largely aligned during the training of elastic networks using the stiffness and rest lengths degrees of freedom ($N=512, 1024$, $M_S=10, M_T=3$).
\label{fig:LearnVTune}}
\end{figure}

We now directly compare results of 
the two learning methods in flow and elastic networks. First, we use both methods to train flow networks of size $N=512$ nodes on the tasks described in section IIa (Fig.~\ref{fig:LearnVTune}a). We find that on average, coupled learning is as efficient as tuning (gradient descent) for training these networks. Indeed, for the problem shown shown in Fig.~\ref{fig:TrainFlow}d, coupled learning actually converges faster than tuning.

To explore the comparison further, we denote the conductance modification prescribed by coupled learning as $\dot{\bm{k}}_{CL}$ (Eq.~\ref{eq:FlowNetworkRule}) and the conductance modification dictated by global supervised learning, or tuning, as $\dot{\bm{k}}_G$ (Eqs.~\ref{eq:CostFunctionMain}-\ref{eq:CostGradientMain}). We argue theoretically in Appendix B that coupled learning leads to modifications to the learning degrees of freedom $k_j$ that are similar to those obtained by tuning using gradient descent (Eq.~\ref{eq:CostGradientMain}). Therefore, we compare the two modification vectors directly by taking a dot product between them (after normalization):

\begin{equation}
\begin{aligned}
&\bm{\dot{k}}_{CL}\sim \partial_{\bm{k}} \{\mathcal{P}^{F}-\mathcal{P}^{C}\} \\
&\bm{\dot{k}}_{G} \sim -\partial_{\bm{k}} \mathcal{C} .
\label{eq:TuningVSLearning}
 \end{aligned}
\end{equation}

These two modification vectors are compared in Fig.~\ref{fig:LearnVTune}(b-d) for flow networks with $N=512$ nodes, trained with $10$ random sources and varying numbers of random targets. Most notably, we observe that the two vectors are indeed nearly identical (the dot product of the normalized vectors is close to unity) for sufficiently small nudges $\eta$ (Fig.~\ref{fig:LearnVTune}c). This is remarkable as the vectors reside in the very high ($\sim 1000$) dimensional space of conductance values $\bm{k}$ of the network. Thus Fig.~\ref{fig:LearnVTune} shows that the local process of coupled learning follows a trajectory in conductance space that is very similar to the trajectory followed during rational tuning, which is a global supervised learning process (although we do note fluctuations in the dot product, see the zoomed scale in Fig.~\ref{fig:LearnVTune}b and appendix D). For small nudges $\eta \lesssim 0.1$, we observe a plateau in this dot product, so the precise choice of $\eta$ is not important (choosing small $\eta\rightarrow 0$ leads to a well-defined limit of the coupled learning rule of Eq.~\ref{eq:TuningVSLearning}; see Appendix B). On the other hand, choosing large $\eta\sim 1$ (as in Hebbian contrastive learning) leads to significant differences between the two processes. We also note that for more complicated functions (more targets to satisfy), the dot product slightly decreases. This effect is exhibited in Fig.~\ref{fig:LearnVTune}(d) for networks of two sizes. However, even with extra functional complexity, coupled learning yields results that are remarkably similar to tuning.


Unlike flow networks, elastic spring networks are fundamentally nonlinear, so it is interesting to explore whether coupled learning produces similar modifications to the learning degrees of freedom as tuning (gradient descent). In Fig.~\ref{fig:LearnVTune}(e-f), we show the alignment between the coupled learning and tuning. We find that the alignment between these modification vectors is still very significant, especially for training with the stiffness degrees of freedom $k_j$. We further find that the alignment is reduced during training as the network approaches low-error solutions, in particular for training the rest lengths $\ell_j$. While this result is interesting and suggests that coupled learning may differ significantly from tuning in certain nonlinear networks, we note that training the network is successful regardless of the deviation between the methods. We leave more detailed exploration of the alignment between possible local learning rules to future study.


A physical learning process called ``directed aging" has recently been introduced to create mechanical metamaterials from mechanical networks~\cite{pashine2019directed,hexner2019effect,hexner2019periodic} and self-folding sheets~\cite{stern2020supervised}. The strategy exploits the idea that stressed bonds in elastic networks tend to weaken (age) over time. During training, the network is nudged in a similar fashion to the coupled learning process. Refs.~\cite{pashine2019directed,hexner2019effect} consider two different classes of learning degrees of freedom in central-force spring networks. In one case, the ``k-model," the stiffnesses $k_j$ of the springs are modified, while in the ``$\ell$-model" rest lengths $\ell_{0,j}$ are modified. The clamped state can either correspond to Eq.~\ref{eq:ClampeddefG} with $\eta=1$ or to a periodically varying amplitude $\eta=\sin{\Omega t}$. In the latter case, the oscillation is performed quasistatically in the sense that the physical degrees of freedom are relaxed completely at each time step. 

In elastic networks, the directed aging $k$-model and $\ell$-model learning rules are:

\begin{equation}
\begin{aligned}
\dot{k_j} = &-\frac{\alpha}{2} [r_j^{C}-\ell_j]^2\, \\
\dot{\ell_j} =& -\alpha k_j [r_j^{C}-\ell_j].
\end{aligned}
  \label{eq:ElasticDirectedAging}
\end{equation}
where the learning degrees of freedom (stiffness or rest length) evolve in response to the clamped boundary conditions.
Such dynamics have the effect of lowering the elastic energy of the desired state, and thus the response of the network to the specified inputs is expected to improve. Indeed, directed aging was shown to be successful in training elastic networks with nearly marginal coordination, so that they lie just above the minimum number of edges per node ($Z_c=2d$) required for mechanical stability in $d$ dimensions. 
Note that because the clamped physical cost function $E^{C}$ can be written as a sum over individual costs of edges $j$, the directed aging rule is local 
as it is for coupled learning. Directed aging therefore corresponds to a physical learning rule, as has been demonstrated experimentally~\cite{pashine2019directed}.

However, directed aging fails to to achieve the desired target response in many instances. In particular, it is observed that either flow networks or highly coordinated elastic networks cannot be trained by directed aging to perform allosteric functions. Comparing directed aging (Eq.~\ref{eq:ElasticDirectedAging}) with coupled learning (Eq.~\ref{eq:ElasticNetworkRule}), we see that the clamped terms are the same, 
but the directed aging rule is missing the free term. 
As a special case of coupled learning, directed aging is expected to perform well in systems whose energy 
in the free state (or its derivative) 
can be neglected. Appendix B shows that in general both the free and clamped terms are necessary to train networks for desired functions. Therefore coupled learning can be viewed as a generalization of the directed aging framework that is successful for a more general class of physical networks.


\section{Discussion}

In this work we have introduced coupled learning, a class of learning rules born of contrastive Hebbian learning and equilibrium propagation~\cite{movellan1991contrastive,scellier2017equilibrium}, and applied it to two types of physical systems, namely flow networks and mechanical spring networks. The advantage of such supervised learning rules, compared to more traditional techniques such as optimization of a cost function, is that they are physically plausible; at least in principle, coupled learning can be implemented in realistic materials and networks, allowing them to learn autonomously from external signals.

Such learning machines are not only interesting in themselves, but may have important advantages compared to physical systems designed by optimizing a cost function. First, because the process involves local responses to local stresses, the approach is scalable--training steps in networks of different sizes (different numbers of nodes) would take approximately the same amount of time. In contrast, the time required to compute gradients of a collective cost function increases rapidly with system size.

Second, the ability to train the system \emph{in-situ} means that it is not necessary to know the network geometry or topology, or even any information about the physical or learning degrees of freedom. This is particularly valuable for experimental systems, which do not have to be characterized in full microscopic detail in order to be trained, as long as the proper learning rules can be implemented, at least in an approximate form. 

Third, as long as the learning rules can be implemented, one does not need to manipulate individual edges to have the desired values of the learning degree of freedom (e.g. the edge conductance for a flow network). Thus, the role of the supervisor can be filled by an end user, training the network for their desired tasks, rather than an expert designer.

An experimental realization of a learning flow network seems quite plausible, as has also been suggested for analog resistor networks~\cite{kendall2020training}. The required ingredients are pipes whose conductances can be modified in response to the current carried by the pipe. It is possible that this condition is similar to that used by the brain vasculature, where vessels can be expanded or constricted~\cite{cipolla2009cerebral, gao2015mechanical}.


We have focused primarily on training physical networks for one particular function (i.e. one particular source-target map, namely allostery). However, coupled learning rules may be used as a stochastic gradient descent step, training the network for a different function in each training iteration. This idea allowed us to train the flow network to distinguish MNIST digits. The dynamics of learning multiple functions using coupled learning is quite involved, and the training performance may depend strongly on the order and frequency of shown training examples. We will address these issues in subsequent work.

One might ask whether physical networks could compete as learning machines with computational neural networks. Our aim is not to outperform neural networks. Rather, the goal is to design physical systems capable of adapting autonomously to tasks and able to generalize to diverse inputs without a computer. Nevertheless, physical networks do supply at least one potential advantage compared to computational neural networks. In contrast to feed-forward neural networks often used in machine learning, the input-output relations in our physical (recurrent) networks are necessarily symmetric in the linear response regime (the regime in which the target response is proportional to the source). \added{As a result, training target responses for given sources may yield a \emph{generative model}~\cite{ng2002discriminative,jebara2012machine}. Such generative models could produce examples of a class given its label by imposing target values distributed around the trained responses, and reading out the free source values. We leave the prospects of physical learning of generative models to future study.}

\subsection*{Acknowledgments}
We thank Nidhi Pashine, Sidney Nagel, Ben Lansdell, Konrad Kording, Vijay Balasubramanian, Eleni Katifori, Douglas Durian and Sam Dillavou for insightful discussions. This research was supported by the U.S. Department of Energy, Office of Basic Energy Sciences, Division of Materials Sciences and Engineering under Awards DE- FG02-05ER46199 and DE-SC0020963 (MS), and the Simons Foundation for the collaboration ``Cracking the Glass Problem'' award $\#$454945 to AJL, as well as Investigator award $\#$327939 (AJL). DH wished to thank the Israel Science Foundation (grant 2385/20).


\appendix

\section{Nudge amplitude $\eta$}

In the main text we used local learning rules based on the difference between the free state and a slightly nudged clamped state ($\eta\ll 1$). We showed how choosing $\eta\ll 1$ in flow networks allows the learning rule to mimic the optimization of a global cost function (Fig.~\ref{fig:LearnVTune}a), and discuss that further in Appendix B.

Here we argue that choosing large nudge amplitudes $\eta \sim 1$, as suggested by contrastive Hebbian learning~\cite{movellan1991contrastive}, can adversely affect learning in physical networks, particularly in non-linear networks (e.g. mechanical networks). 

The choice $\eta \ll 1$ implies that the clamped state is almost identical to the free state. Inspecting Eq.~\ref{eq:FlowNetworkRule} or Eq.~\ref{eq:ElasticNetworkRule}, we see that the learning rule essentially becomes a derivative of the energy function with respect to the physical variables, in the direction towards the desired state. Thus, the learning rule is in fact a derivative of the energy in both spaces, those of the physical (e.g. pressure values) and learning degrees of freedom (e.g. pipe conductances). The choice $\eta \ll 1$ implies a \textit{local} modification of the system not only in the spatial sense (so that only nearby elements communicate), but also in the generalized energy landscape of the combined configuration space of physical and learning parameters.

Conversely, choosing a large nudge amplitude $\eta \sim 1$ means the free and clamped states can be far away from each other, so that the learning rule does not approximate the derivative very well. This may be particularly important in non-linear networks whose energy landscape is non-convex. In such cases, if the free and clamped states are far apart, they may belong to different attractor basins in the landscape, possibly impeding the learning process, since it is not guaranteed that the learning rule can eliminate energy barriers between the two states. This problem has been long recognized as a limitation of contrastive Hebbian learning~\cite{movellan1991contrastive}, and has been solved by using a small nudging factor~\cite{scellier2017equilibrium}.

\begin{figure}
\includegraphics[width=1.0\linewidth]{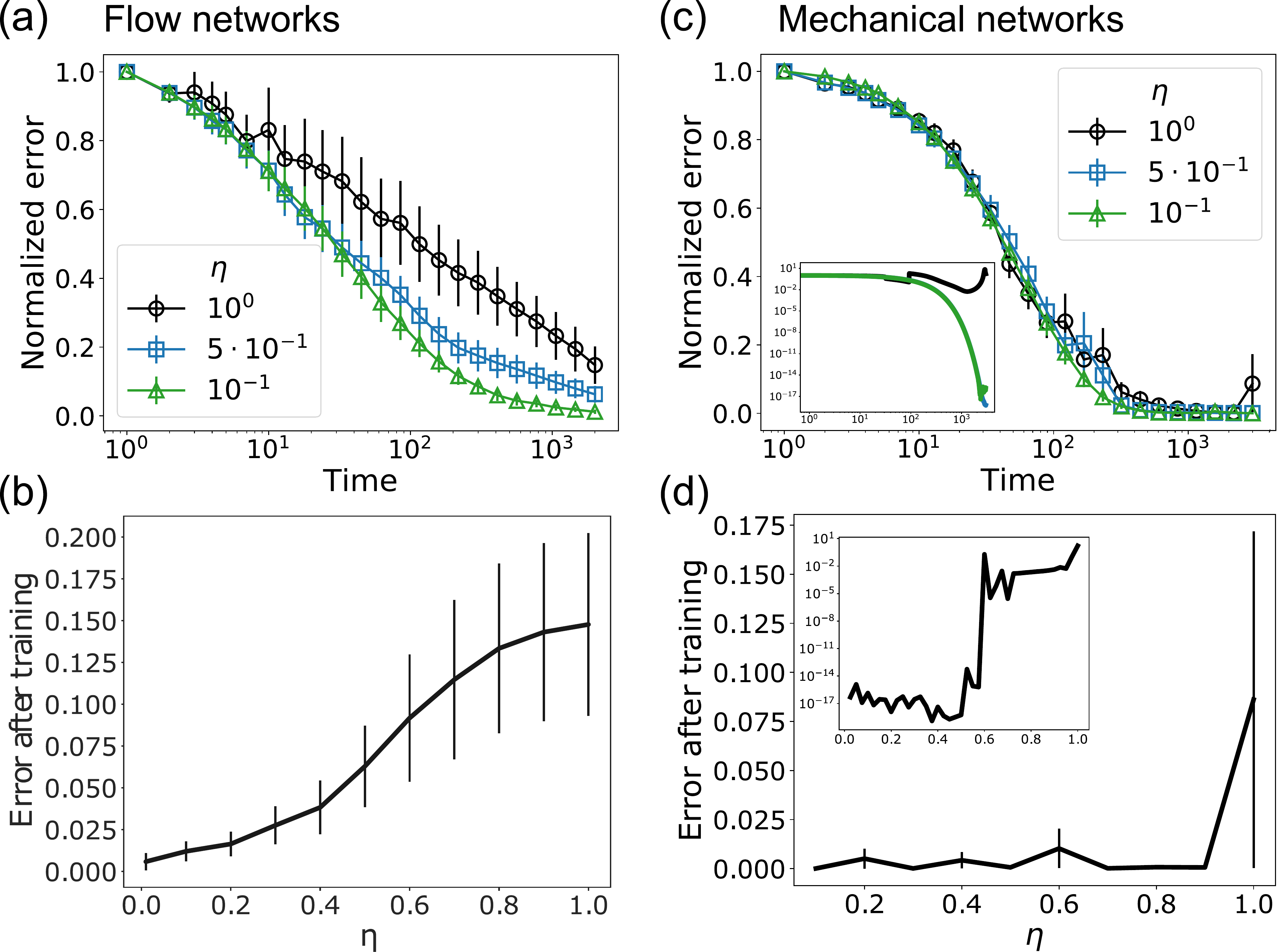}
\caption{Effects of nudge amplitude $\eta$ on learning. a) $512$ node flow networks trained with $10$ random source nodes and $3$ target nodes. As the nudge amplitude $\eta$ is increased, training gradually and continuously becomes slower. b) Trained flow networks (after $2\cdot 10^3$ time steps) perform better if trained with small nudges. c) When learning is attempted in non-linear mechanical networks using the $\ell$-rule, a similar picture emerges \textit{on average}, where learning degrades at higher eta. Moreover, at particular realizations, we observe non-continuous degradation of the learning process at particular values of $\eta$ (inset). d) On average, learning performance is largely maintained until $\eta \sim 1$. In one $512$ node network, trained with $10$ random sources and $3$ random targets, training is relatively stable until $\eta\approx 0.6$, and then goes thorough a discontinuous jump (inset). Averaged data results from 17 realization of learning problems.
\label{fig:Eta}}
\end{figure}

We thus argue for the benefit of choosing small nudges $\eta \ll 1$. To test this proposal, we observe the learning process in flow networks and mechanical networks with different choices of $\eta$ (Fig.~\ref{fig:Eta}). We trained flow networks (with $512$ nodes, $10$ sources and $3$ targets) with varying values of the nudge parameter $\eta$ (Fig.~\ref{fig:Eta}ab). It is generally found that choosing small nudge values $\eta\ll 1$ leads to better learning performance. The learning process is faster for lower $\eta$, so that after a fixed training time, the accuracy of networks trained with small $\eta$ is better. When $\eta$ is raised gradually, we find that the learning process is gradually and continuously slowed down.

Changing the nudge amplitude for non-linear networks, such as a mechanical spring network (Fig.~\ref{fig:Eta}cd) can lead to qualitatively different results. In one network of $512$ nodes (trained with the $\ell$-learning rule of Eq.~\ref{eq:ElasticNetworkRule}, $10$ random sources and $3$ random targets), we find a discontinuous behavior at $\eta\approx 0.6$. At lower values we observe similar behavior to a linear flow network, but at higher $\eta$ learning is abruptly slowed down by orders of magnitude. As discussed above, this effect likely stems from the issue that for this learning problem, the initial free state and the desired states belong to different attractor basins in the network energy landscape. Thus, when $\eta$ is large enough, the free and clamped states belong to different basins, significantly slowing down the learning process. That being said, these nonlinear networks can still learn with large nudges $\eta \sim 1$, reaching low error values. 

We conclude that choosing small nudge values $\eta \ll 1$ yields superior results when training our networks. Such choice leads to consistent and continuous learning in both linear and non-linear networks, as the learning rule best approximates the gradient descent on the desired cost function (see Appendix B). When choosing large nudges $\eta\sim 1$, learning efficiency usually deteriorates, and we observe interesting dynamical effects such as intermittent unlearning events. We will address the dynamics of large nudges in future work.

\section{Effective cost function}

In the main text we established that the coupled learning rule trains networks to present desired source-target relations. Surprisingly, the local learning rule predicted by our framework is very similar to the global modification resulting from minimizing a cost function. This result is very encouraging, as it shows that physically plausible, local learning rules can give rise to comparable training as non-local gradient descent, which is the ultimate greedy training strategy. In this section we explain this surprising similarity by deriving an effective cost function from our learning rule.

As discussed in the main text, the network autonomously modifies its edge parameters $\bm{w}$ by subtracting the energies of free and clamped states and taking a derivative:

\begin{equation}
\begin{aligned}
\dot{\bm{w}} \sim \partial_{\bm{w}} \{&E^{F}(x_S,x_T^{F},x_H^{F})\\
-&E^{C}(x_S,x_T^{F}+\eta [X_T- x_T^{F}],x_H^{C})\}.
\end{aligned}
  \label{eq:SIGeneralRule}
\end{equation}

Here, we expanded the notation of the main text to include the free and clamped states of all the hidden nodes $x_H$. Given a small nudge parameter $\eta\ll 1$, the clamped state is very similar to the free state, and we can approximate the clamped state using a Taylor expansion around the free state. In particular, let us write the clamped state of the hidden nodes as $x_H^{C}\equiv x_H^{F}+ \Delta x_H$, with the shift is $\Delta x_H$ a vector of magnitude $\sim\eta$. Further, we define the signed error in the target signal $\epsilon_T \equiv X_T- x_T^{F}$. Concatenating the source, target and hidden nodes together, the difference between the free and clamped state can be written as 
$$\Delta x \equiv (x_S^C-x_S^F , x_T^C-x_T^F , x_H^C-x_H^F)=\eta (0,\epsilon_T,\Delta x_H).$$ Now, we can approximate the energy of the clamped state $E^{C}$ as

\begin{equation}
\begin{aligned}
E^{C}\approx &E^{F} +\eta (0,\epsilon_T,\Delta x_H) \nabla_x E^{F}+\\
&+\frac{1}{2}\eta^2 (0,\epsilon_T,\Delta x_H)^T \frac{\partial^2 E^{F}}{\partial x \partial x}(0,\epsilon_T,\Delta x_H)
\end{aligned}
  \label{eq:TaylorApprox}
\end{equation}

By definition of the free state, it is a stable equilibrium state of the physical network, so that the first order term vanishes $\nabla_x E^{F}=0$. The second derivative in the second order term is the Hessian at the free state $H^{F}$. Thus the Taylor expansion simplifies to

\begin{equation}
\begin{aligned}
E^{C}\approx E^{F}+\frac{\eta^2}{2} (0,\epsilon_T,\Delta x_H)^T H^{F}(0,\epsilon_T,\Delta x_H).
\end{aligned}
  \label{eq:TaylorSimplifies}
\end{equation}

This result can be plugged back into the learning rule of Eq.~\ref{eq:SIGeneralRule} to obtain

\begin{equation}
\begin{aligned}
\dot{\bm{w}} \sim - \frac{\eta^2}{2} \partial_{\bm{w}} (0,\epsilon_T,\Delta x_H)^T H^{F}(0,\epsilon_T,\Delta x_H).
\end{aligned}
  \label{eq:SIGeneralRule2}
\end{equation}

As discussed in the main text, a typical tuning algorithm starts by defining a cost function $\mathcal{C}\equiv \epsilon_T^2=\epsilon_T^T I \epsilon_T$, with $I$ the identity matrix. This cost function is minimized via gradient descent

\begin{equation}
\begin{aligned}
\dot{\bm{w}}_{G} \sim -\partial_{\bm{w}} \mathcal{C}.
\end{aligned}
  \label{eq:SITuning}
\end{equation}

Comparing equations~\ref{eq:SIGeneralRule2} and~\ref{eq:SITuning}, we may identify an effective cost function minimized by our learning rule

\begin{equation}
\begin{aligned}
\mathcal{C}&\sim \epsilon_T^T I \epsilon_T\\
\mathcal{C}_{\textrm{eff}} &\sim (0,\epsilon_T,\Delta x_H)^T H^{F}(0,\epsilon_T,\Delta x_H).
\end{aligned}
  \label{eq:SIEffectiveCost}
\end{equation}

While the effective cost function we derived is not identical to the standard cost function, they share important properties. Firstly, both functions are convex (square forms) in the difference between the desired and the obtained result $\epsilon_T$. Both functions are bounded from below by $0$, as the Hessian of the stable free state is positive semi-definite. The two cost functions vanish if and only if the obtained targets are equal to the desired targets $x_T^{F}=X_T$. Note that this is true since at $\epsilon_T=0$, $\Delta x_H$ vanishes as well. Overall, we know that a global minimum of $\mathcal{C}_{\textrm{eff}}$ must be a global minimum of $\mathcal{C}$. The effect of the Hessian matrix close to that minimum is mostly to stretch and rotate the paraboloid cup implied by the square form. When comparing the two cost functions, we also find that their gradients are aligned to significant extents as shown in Fig.~\ref{fig:LearnVTune}. While this is alignment is not guaranteed and affected by many factors (e.g. network physics and architecture, number of targets, error values, etc.), the fact that both cost functions share the same minima means that coupled learning will tend to find solutions that minimize the original cost function too. We conclude that minimizing $\mathcal{C}_{\textrm{eff}}$ should mimic the minimization of $\mathcal{C}$, at least close enough to a minimum of $\mathcal{C}$, as is in fact observed.

Recently, it was shown~\cite{kendall2020training, laborieux2020scaling} that using a nudged state that applies a `force' on the output nodes, rather than clamping their values as done here, implies that the effective cost function approaches the original cost function $\mathcal{C}$ for small nudge amplitudes $\eta\rightarrow 0$. 

\section{Experimentally feasible learning rule for flow networks}

As discussed in the main text, implementing the full coupled learning rule of Eq.~\ref{eq:FlowNetworkRule} in an experimental setting may prove difficult. One would need to implement pipes that can adapt in opposite ways to the same flows, depending if the system is set in the free or clamped state. To simplify this learning rule for better experimental feasibility, we start by considering the full rule for conductance adaptation

\begin{equation}
\begin{aligned}
\dot{\bm{k}} \sim \{&[\bm{\Delta p}^{F}(p_S,p_T^{F})]^2\\ 
- & [\bm{\Delta p}^{C}(p_S,p_T^{F}+\eta [P_T- p_T^{F}])]^2\}.
\end{aligned}
  \label{eq:SIFlowNetworkRule}
\end{equation}

Recall that $\bm{\Delta p}$ is a vector of pressure differences across all pipes. It may be replaced by the difference of pressures in the two nodes connected by the pipe $\bm{\Delta p}\equiv \bm{p}_2-\bm{p}_1$. We can use this form in Eq.~\ref{eq:SIFlowNetworkRule} to obtain:

\begin{equation}
\begin{aligned}
\dot{\bm{k}} &\sim \{[\bm{p}_2^{F}-\bm{p}_1^{F}]^2- [\bm{p}_2^{C}-\bm{p}_1^{C}]^2\}=\\
&= (\bm{p}_1^{F}-\bm{p}_1^{C})(\bm{p}_1^{F}+\bm{p}_1^{C}) + 2\bm{p}_1^{C}\bm{p}_2^{C} + \\
&+(\bm{p}_2^{F}-\bm{p}_2^{C})(\bm{p}_2^{F}+\bm{p}_2^{C}) -2\bm{p}_1^{F}\bm{p}_2^{F}
\end{aligned}
  \label{eq:SIFlowNetworkRule2}
\end{equation}

Now, we define the difference between the free and clamped states as $\bm{\delta p_\mu} \equiv \bm{p_\mu^{F}}-\bm{p_\mu^{C}}$, with $\mu$ the node index. If we choose a small nudge parameter $\eta\ll 1$, the free and clamped states are almost identical, such that $\bm{p_\mu^{F}}\approx \bm{p_\mu^{C}}\rightarrow \vert \bm{\delta p_\mu} \vert \ll \vert \bm{p_\mu^{F}} \vert$. Using the proximity of the free and clamped states, we can approximate Eq.~\ref{eq:SIFlowNetworkRule2}

\begin{equation}
\begin{aligned}
\dot{\bm{k}}\sim 2[&\bm{\delta p_1}\bm{p_1^{F}}+\bm{\delta p_2}\bm{p_2^{F}}-\\
-&\bm{p_1^{F}}\bm{p_2^{F}}+(\bm{p_1^{F}}-\bm{\delta p_1})(\bm{p_2^{F}}-\bm{\delta p_2})]\approx\\
\approx  2 &(\bm{\delta p_1}-\bm{\delta p_2})(\bm{p_1^{F}}-\bm{p_2^{F}}).
\end{aligned}
  \label{eq:SIFlowNetworkRule3}
\end{equation}

Going back to the notation of pressure drops across pipes, we can write $\bm{\Delta p}^{\delta}\equiv \bm{\delta p_1}-\bm{\delta p_2}$, supporting the definition of the $\delta$-state as described in the main text. The approximated learning rule can thus be rewritten as 
\begin{equation}
\begin{aligned}
\dot{\bm{k}}\approx 2\alpha\eta^{-1} \bm{\Delta p}^{F} \bm{\Delta p}^{\delta}.
\end{aligned}
  \label{eq:SIFlowNetworkRuleApprox}
\end{equation}

As long as we pick $\eta \ll 1$, this approximation is very accurate. To make this learning rule more experimentally feasible, we further approximate it so that only the sign of $\bm{\Delta p}^{F}$ is accounted for, and the conductances change in only one direction, yielding Eq.~\ref{eq:DeltaRule}. Finally, we note that while this approximation is derived specifically for linear flow networks, the derivation may be similar for diverse physical networks, as long as the energy of an edge is (to first approximation) proportional to the difference in `activation' between the connected nodes. For example, similar approximation could be derived for both the $k$ and $\ell$ learning rules in elastic networks (Eq.~\ref{eq:ElasticNetworkRule}).

\section{Non-negative learning parameters}

\begin{figure}
\includegraphics[width=0.98\linewidth]{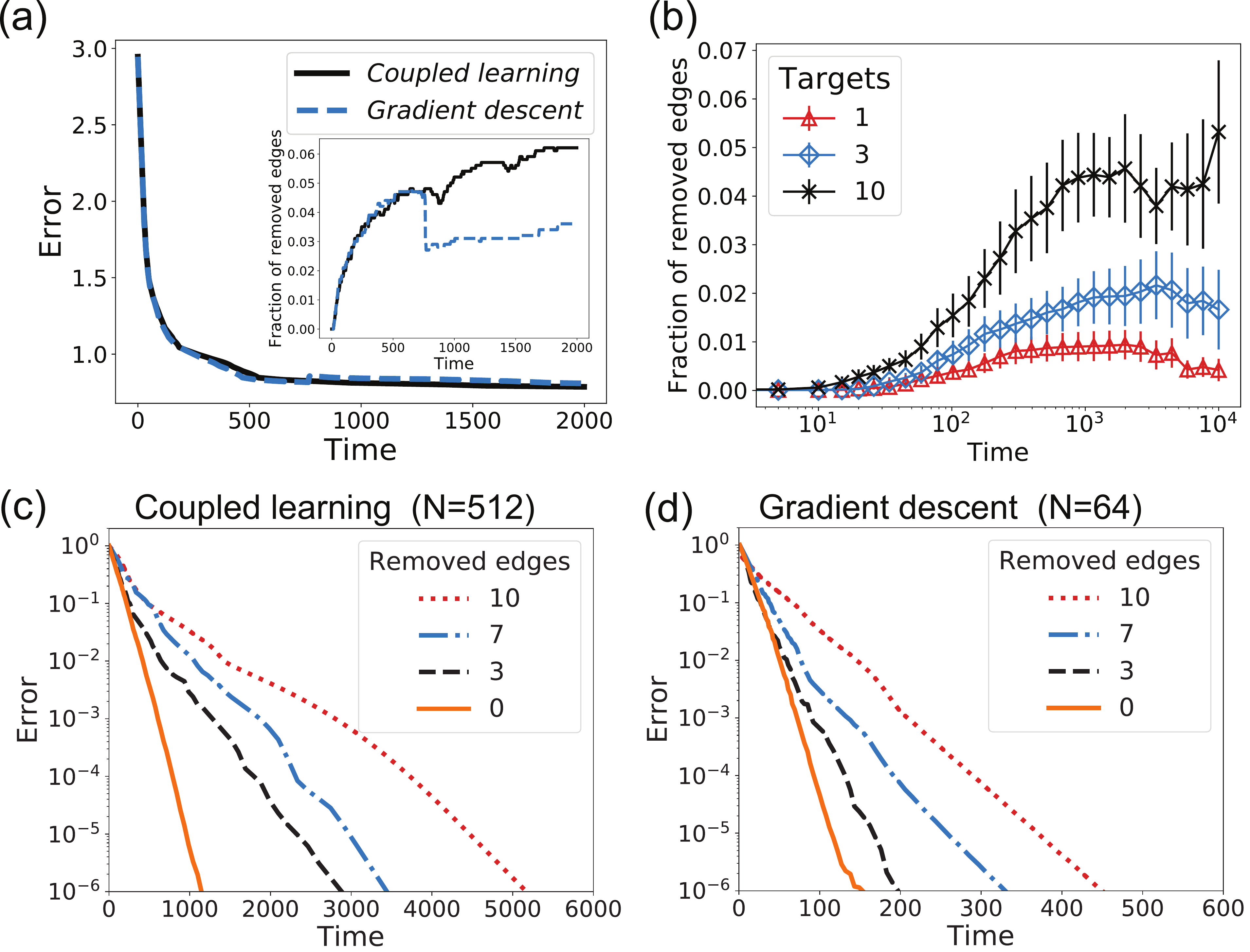}
\caption{Effects of non-negative learning degrees of freedom. a) Training a flow network ($N=512, M_S=10, M_T=10$) on a `hard' task where learning only reduces the error by a small factor. We find that when coupled learning fails to train a flow network, gradient descent (tuning) fails as well, suggesting the issue lies with the network and not the learning rule. Inset shows that a few percent of edges (conductances) reach their minimal value during training. b) Averaging over $10$ training tasks, we find that harder training problems with more simultaneous targets induce a larger fraction of edges to reach their minimal value. c) \added{Median normalized error during training with coupled learning for networks with different number of vanishing edges, showing that the non-negative constraint slows learning ($N=512, M_S=10, M_T=5$).} d) \added{A similar slow-down arises when the network is trained with standard gradient descent ($N=64, M_S=10, M_T=5$).} 
\label{fig:PosK}}
\end{figure}

In this work we introduced coupled learning, a supervised learning framework for physical networks. Using coupled learning, one can derive the relevant learning rules (i.e. the proper dynamics for modifying the learning degrees of freedom). The systems demonstrated in this work, flow and elastic networks, are special examples. However, the learning degrees of freedom in both examples have a shared physical limitation: The conductance of pipes in a flow network, as well as the stiffness and rest lengths of springs in an elastic networks, are all defined as non-negative quantities. This is a similar issue to the non-negativity of synaptic connection strengths in natural neural networks~\cite{parisien2008solving}. 

Due to this physical limitation of the networks we considered, their capacity to learn arbitrary mappings between the sources and targets is limited. For example, while linear flow networks will always give rise to a linear mapping between the pressures at the sources and targets, the non-negativity limitation excludes many conceivable linear mappings, depending on the geometry of the network and source pressures. Therefore, we expect that a linear flow network would generally be less successful than a general linear model (i.e. linear regression) in learning certain desired tasks.

While several possible approaches for circumventing this issue have been proposed~\cite{hu2016dot,wang2019reinforcement}, here we discuss the effect of such non-negativity limitation on the learning process in our physical networks. To avoid negative values in the learning degrees of freedom, we simply cut-off their values at a small number (e.g. $k\geq 10^{-6}$ in flow networks). We find that training the networks often causes a fraction of the learning parameters to reach these small cut-off values. This may significant affect learning for two reasons. First, the magnitude of modification to the learning parameters is reduced, so that learning may fail or slow down. Second, the modification of the learning parameters becomes less aligned with the gradient of the cost function. 

In Fig.~\ref{fig:PosK}, we again directly compare coupled learning and tuning (gradient descent) for training of flow networks, as shown in Fig.~\ref{fig:LearnVTune}. However, in this example ($N=512, M_S=10, M_T=10$), learning `fails', so that the initial error is only reduced by a factor of $3$. It is important to note that the failure is shared by both coupled learning and gradient descent, hinting that the problem is not in the learning rule, but rather in the network physics or architecture. Indeed, Fig.~\ref{fig:PosK}a inset shows that as training proceeds, a few percent of the edges vanish, with their conductance values reaching the minimal cutoff. In Fig.~\ref{fig:PosK}b, we average the fraction of vanishing edges over $10$ training examples to show that harder training problems (with more simultaneous targets) often lead to a larger fraction of vanishing conductance values. One may ask whether this vanishing of edges causes problems for training. \added{In Fig.~\ref{fig:PosK}cd, we find that learning slows down considerably with an increasing number of removed edges for tasks with the same number of targets ($M_T=5$), regardless if training is performed using our coupled learning approach or tuning (gradient descent). Note that the gradient descent case trains faster only because we consider a much smaller $N=64$ network (see Fig.~\ref{fig:TrainFlow}ef). Together with Fig.~9b, these results suggest that it is the physical constraint of non-negative learning degrees of freedom that slows down learning, not the coupled learning protocol itself.}



\bibliographystyle{unsrt}
\bibliography{Citations}

\end{document}